# Highly Parallel Sparse Matrix-Matrix Multiplication☆,☆☆

Aydın Buluç*,[1]

*High Performance Computing Research, Lawrence Berkeley National Laboratory, 1 Cyclotron Road, Berkeley, CA 94720*

John R. Gilbert

*Computer Science Department, University of California, Santa Barbara, CA 93106-5110*

---

**Abstract**

Generalized sparse matrix-matrix multiplication is a key primitive for many high performance graph algorithms as well as some linear solvers such as multigrid. We present the first parallel algorithms that achieve increasing speedups for an unbounded number of processors. Our algorithms are based on two-dimensional block distribution of sparse matrices where serial sections use a novel hypersparse kernel for scalability. We give a state-of-the-art MPI implementation of one of our algorithms. Our experiments show scaling up to thousands of processors on a variety of test scenarios.

---

☆Part of the material in this paper previously appeared in preliminary form in the proceedings of the 22th International Parallel and Distributed Processing Symposium (IPDPS'08) and the 37th International Conference on Parallel Processing (ICPP'08)

☆☆This work was supported in part by NSF grant CNS-0709385 and in part by the National Science Foundation through TeraGrid resources provided by TACC under grant number TG-CCR090036

*Corresponding author

*Email addresses:* abuluc@lbl.gov (Aydın Buluç), gilbert@cs.ucsb.edu (John R. Gilbert)

[1]Part of this work is performed while the author was with the University of California, Santa Barbara

## 1. Introduction

Development and implementation of large-scale parallel graph algorithms poses numerous challenges in terms of scalability and productivity [1, 2]. Linear algebra formulations of many graph algorithms already exist in the literature [3, 4, 5]. By exploiting the duality between matrices and graphs, linear algebraic formululations aim to apply the existing knowledge on parallel matrix algorithms to parallel graph algorithms. One of the key linear-algebraic primitives for graph algorithms is computing the product of two sparse matrices (SpGEMM) over a semiring. It serves as a building block for many algorithms including graph contraction [6], breadth-first search from multiple source vertices, peer pressure clustering [7], recursive formulations of all-pairs shortest-paths algorithms [8], matching algorithms [9], and cycle detection [10], as well as for some other applications such as multigrid interpolation/restriction [11], and parsing context-free languages [12].

Most large graphs in applications, such as the WWW graph, finite element meshes, planar graphs, and trees, are sparse. In this work, we consider a graph to be sparse if $nnz = O(n)$, where $nnz$ is the number of edges and $n$ is the number of vertices. Dense matrix multiplication algorithms are inefficient for SpGEMM since they require $O(n^3)$ space and the current fastest dense matrix multiplication algorithm runs in $O(n^{2.38})$ [13, 14] time. Furthermore, fast dense matrix multiplication algorithms operate on a ring instead of a semiring, which makes them unsuitable for many algorithms on general graphs. For example, it is possible to embed the semiring into the ring of integers for the all-pairs shortest-paths problem on unweighted and undirected graphs [14], but the same embedding does not work for weighted or directed graphs [15].

Let $\mathbf{A} \in \mathbb{S}^{m \times n}$ be a sparse rectangular matrix of elements from an arbitrary semiring $\mathbb{S}$. We use $nnz(\mathbf{A})$ to denote the number of nonzero elements in $\mathbf{A}$. When the matrix is clear from context, we drop the parenthesis and simply use $nnz$. For sparse matrix indexing, we use the convenient M ® colon notation, where $\mathbf{A}(:, i)$ denotes the $i$th column, $\mathbf{A}(i, :)$ denotes the $i$th row, and $\mathbf{A}(i, j)$ denotes the element at the $(i, j)$th position of matrix $\mathbf{A}$. For one-dimensional arrays, $\mathbf{a}(i)$ denotes the $i$th component of the array. Sometimes, we abbreviate and use $nnz(j)$ to denote the number of nonzeros elements in the $j$th column of the matrix in context. Array indices are 1-based throughout this paper. We use flops($\mathbf{A}\, op\, \mathbf{B}$), pronounced "flops", to denote the number of nonzero arithmetic operations required by the operation $\mathbf{A}\, op\, \mathbf{B}$. Again, when the operation and the operands are clear from context, we simply use flops.

The most widely used data structures for sparse matrices are the Compressed Sparse Columns (CSC) and Compressed Sparse Rows (CSR) [16]. The second chapter of the first author's thesis [17] give concise descriptions of common SpGEMM algorithms operating both on CSC/CSR and triples. The SpGEMM problem was recently reconsidered by Yuster and Zwick [18] over a ring, where the authors use a fast dense matrix multiplication such as arithmetic progression [13] as a subroutine. Their algorithm uses $O(nnz^{0.7}\, n^{1.2} + n^{2+o(1)})$ arithmetic operations, which is theoretically close to optimal only if we assume that the number of nonzeros in the resulting matrix $\mathbf{C}$ is $\Theta(n^2)$. This assumption rarely holds in reality. Instead, we provide a work sensitive analysis by expressing the computation complexity of our SpGEMM algorithms in terms of flops.

Practical sparse algorithms have been proposed by different researchers over the years [19, 20] using various data structures. Although they achieve reasonable performance on some classes of matrices, none of these algorithms outperforms the classical sparse matrix-matrix multiplication algorithm for general sparse matrices, which was first described by Gustavson [21] and was used in Matlab [22] and CSparse [23]. The classical algorithm runs in $O(\text{flops} + nnz + n)$ time.

In Section 2, we present two novel algorithms for sequential SpGEMM. The first one is geared towards computing the product of two ***hypersparse*** matrices. A matrix is hypersparse if the ratio of nonzeros to its dimension is asymptotically 0. It is used as the sequential building block of our parallel 2D algorithms described in Section 3. Our H _GEMM algorithm uses a new $O(nnz)$ data structure, called ***DCSC*** for ***doubly compressed sparse columns***, which is explained in Section 2.2. The H - _GEMM is based on the outer-product formulation and has time complexity $O(nzc(\mathbf{A}) + nzr(\mathbf{B}) + \text{flops} \cdot \lg ni)$, where $nzc(\mathbf{A})$ is the number of columns of $\mathbf{A}$ that contain at least one nonzero, $nzr(\mathbf{B})$ is the number of rows of $\mathbf{B}$ that contain at least one nonzero, and $ni$ is the number of indices $i$ for which $\mathbf{A}(:, i) \neq \emptyset$ and $\mathbf{B}(i, :) \neq \emptyset$. The overall space complexity of our algorithm is only $O(nnz(\mathbf{A}) + nnz(\mathbf{B}) + nnz(\mathbf{C}))$. Notice that the time complexity of our algorithm does not depend on $n$, and the space complexity does not depend on flops.

Section 3 presents parallel algorithms for SpGEMM. We propose novel algorithms based on 2D block decomposition of data in addition to giving the complete description of an existing 1D algorithm. To the best of our knowledge, parallel algorithms using a 2D block decomposition have not earlier been developed for sparse matrix-matrix multiplication.

Toledo et al. [24] proved that 2D dense matrix multiplication algorithms are optimal with respect to the communication volume, making 2D sparse algorithms likely to be more scalable than their 1D counterparts. In Section 4, we show that this intuition is indeed correct by providing a theoretical analysis of the parallel performance of 1D and 2D algorithms.

In Section 5, we model the speedup of parallel SpGEMM algorithms using realistic simulations and projections. Our results show that existing 1D algorithms are not scalable to thousands of processors. By contrast, 2D algorithms have the potential for scaling up indefinitely, albeit with decreasing parallel efficiency, which is defined as the ratio of speedup to the number of processors.

Section 6 describes the experimental setup we used for evaluating our Sparse SUMMA implementation, and presents the final results. We describe other techniques we have used for implementating our parallel algorithms, and their effects on performance in Section 7. Section 8 offers some future directions.

## 2. Sequential Sparse Matrix Multiply

We first analyze different formulations of sparse matrix-matrix multiplication using the layered graph model in Section 2.1. This graph theoretical explanation gives insights on the suitability of the outer-product formulation for multiplying hypersparse matrices H _GEMM. Section 2.2 defines hypersparse matrices and Section 2.3 introduces the DCSC data structure that is suitable to store hypersparse matrices. We present our H _GEMM algorithm in Section 2.2.

### *2.1. Layared graphs for different formulations of SpGEMM*

Matrix multiplication can be organized in many different ways. The inner-product formulation that usually serves as the definition of matrix multiplication is well-known. Given two matrices $\mathbf{A} \in \mathbb{R}^{m \times k}$ and $\mathbf{B} \in \mathbb{R}^{k \times n}$, each element in the product $\mathbf{C} \in \mathbb{R}^{m \times n}$ is computed by the following formula:

$$\mathbf{C}(i, j) = \sum_{l=1}^{k} \mathbf{A}(i, l)\mathbf{B}(l, j). \quad (1)$$

This formulation is rarely useful for multiplying sparse matrices since it requires $\Omega(mn)$ operations regardless of the sparsity of the operands.

We represent the multiplication of two matrices **A** and **B** as a three layered graph, following Cohen [25]. The layers have $m$, $k$ and $n$ vertices, in that order. The first layer of vertices ($U$) represent the rows of **A** and the third layer of vertices ($V$) represent the columns of **B**. The second layer of vertices ($W$) represent the dimension shared between matrices. Every nonzero $\mathbf{A}(i, l) \neq 0$ in the $i$th row of **A** forms an edge $(u_i, w_l)$ between the first and second layers and every nonzero in $\mathbf{B}(l, j) \neq 0$ in the $j$th column of **B** forms an edge $(w_l, v_j)$ between the second and third layers.

We perform different operations on the layered graph depending on the way we formulate the multiplication. In all cases though, the goal is to find pairs of vertices $(u_i, v_j)$ sharing an adjacent vertex $w_k \in W$, and if any pair shares multiple adjacent vertices, to merge their contributions.

Using inner products, we analyze each pair $(u_i, v_j)$ to find the set of vertices in $\widetilde{W_{ij}} \subseteq W = \{w_1, w_2, ..., w_k\}$ that are connected to both $u_i$ amd $v_j$ in the graph shown in Figure 1. The algorithm then accumulates contributions $a_{il} \cdot b_{lj}$ for all $w_l \in \widetilde{W_{ij}}$. The result becomes the value of $\mathbf{C}(i, j)$ in the output. In general this inner-product subgraph is sparse, and a contribution from $w_l$ happens only when both edges $a_{il}$ and $b_{lj}$ exist. However, this sparsity is not exploited using inner products as it needs to examine each $(u_i, v_j)$ pair, even when the set $\widetilde{W_{ij}}$ is empty.

In the outer-product formulation, the product is written as the summation of $k$ rank one matrices:

$$\mathbf{C} = \sum_{l=1}^{k} \mathbf{A}(:, l)\mathbf{B}(l, :). \quad (2)$$

A different subgraph results from this formulation as it is the set of vertices $W$ that represent the shared dimension that play the central role. Note that the edges are traversed in the outward direction from a node $w_i \in W$, as shown in Figure 2. For sufficiently sparse matrices, this formulation may run faster because this traversal is performed only for the vertices in $W$ (size $k$) instead of the inner product traversal that had to be performed for every pair (size $mn$). The problem with outer-product traversal is that it is hard to accumulate the intermediate results into the final matrix.

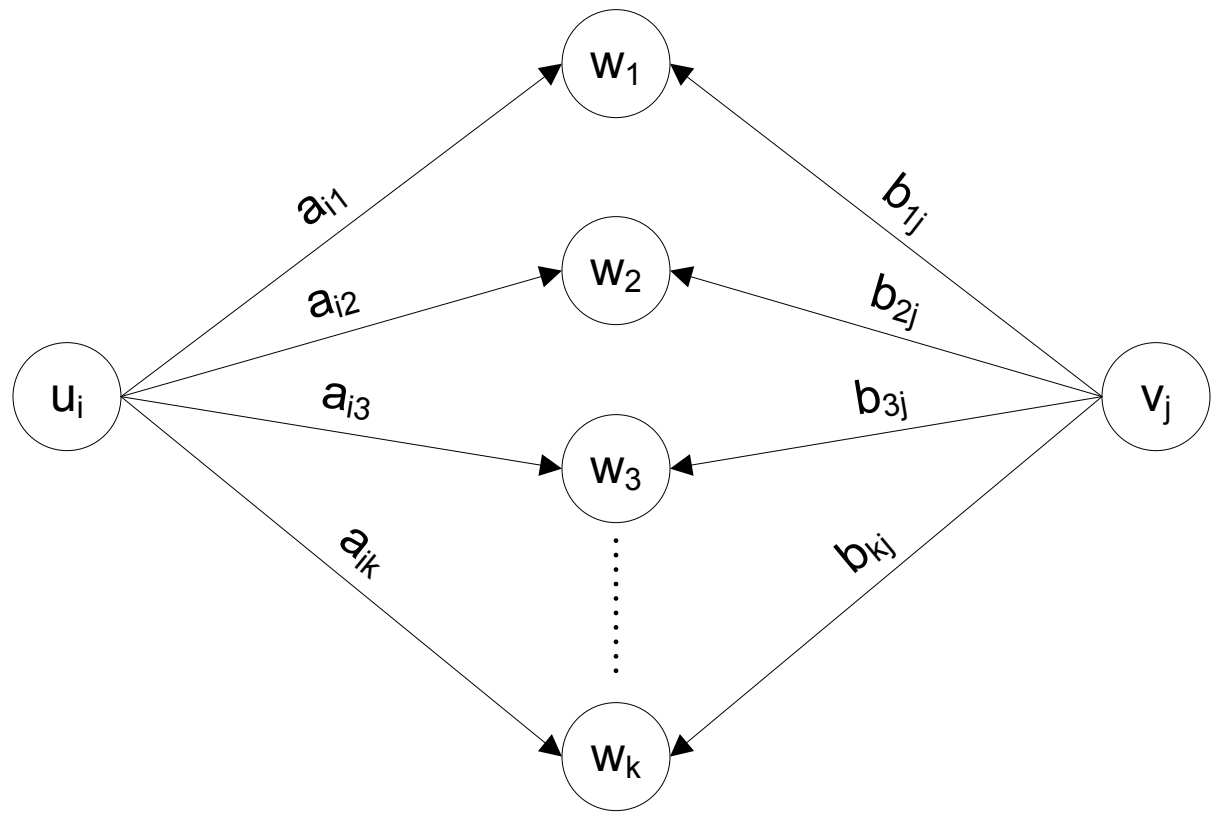

Figure 1: Graph representation of the inner product $\mathbf{A}(i, :) \cdot \mathbf{B}(:, j)$

A row-by-row formulation of matrix multiplication performs a traversal starting from each of the vertices in $U$ towards $V$, as shown in Figure 3 for $u_i$. Each traversal is independent from each other because they generate different rows of $\mathbf{C}$. Finally, a column-by-column formulation creates an isomorphic traversal, in the reverse direction (from $V$ to $U$).

### *2.2. Hypersparse Matrices*

One conventional storage format for sparse matrices is the Compressed Sparse Rows (CSR) format, which stores the nonzeros in consecutive locations and maintains pointers to the first nonzero element of each row. Due to these pointers, it takes $\Theta(n + nnz)$ space for an $n$-by-$n$ matrix. Recall that a matrix is hypersparse if $nnz < n$. Although CSR is a fairly efficient storage scheme for general sparse matrices having $nnz = \Omega(n)$, it is asymptotically suboptimal for hypersparse matrices. Hypersparse matrices are fairly rare in numerical linear algebra (indeed, a nonsingular square matrix must have $nnz \geq n$), but they occur frequently in computations on graphs, particularly in parallel.

Our main motivation for hypersparse matrices comes from parallel processing. Hypersparse matrices arise after the 2-dimensional block data decomposition of ordinary sparse matrices for parallel processing. Consider a sparse matrix with $c$ nonzero elements in each column. After the 2D decomposition of the matrix, each processor locally owns a submatrix with dimensions $(n/\sqrt{p}) \times (n/\sqrt{p})$. Storing each of those submatrices in CSC format takes $\Theta(n\sqrt{p} + nnz)$ space, whereas the amount of space needed to store the whole matrix in CSC format on a single processor is only $\Theta(n+nnz)$. As the number of processors increases, the $n\sqrt{p}$ term dominates the $nnz$ term.

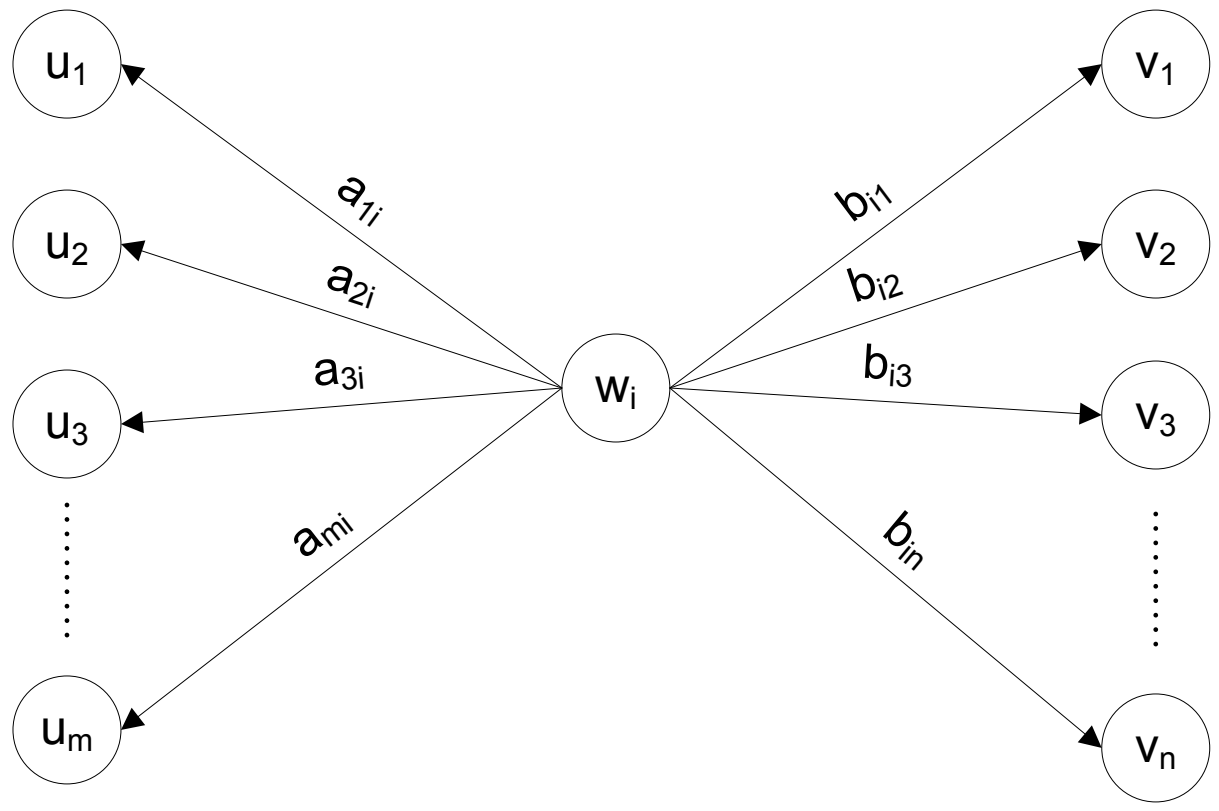

Figure 2: Graph representation of the outer product $\mathbf{A}(:, i) \cdot \mathbf{B}(i, :)$

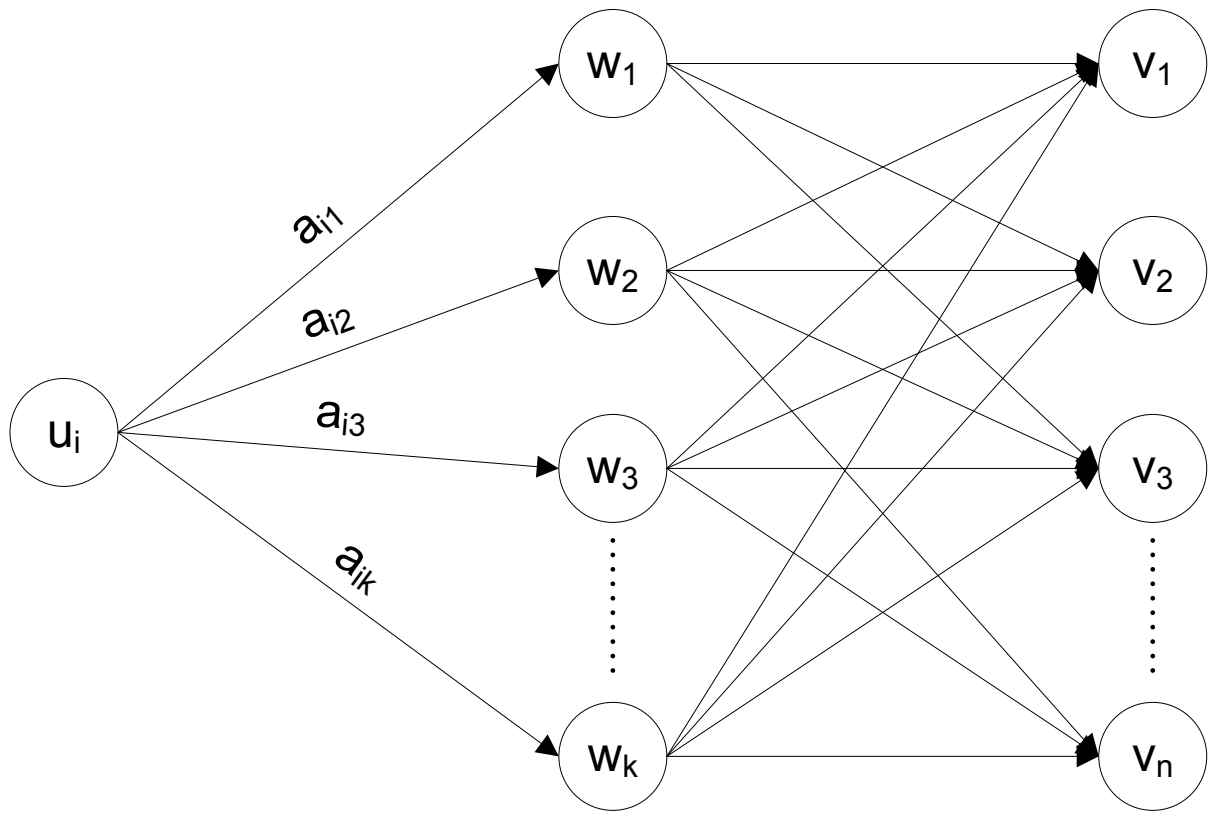

Figure 3: Graph representation of the sparse row times matrix product $\mathbf{A}(i, :) \cdot \mathbf{B}$

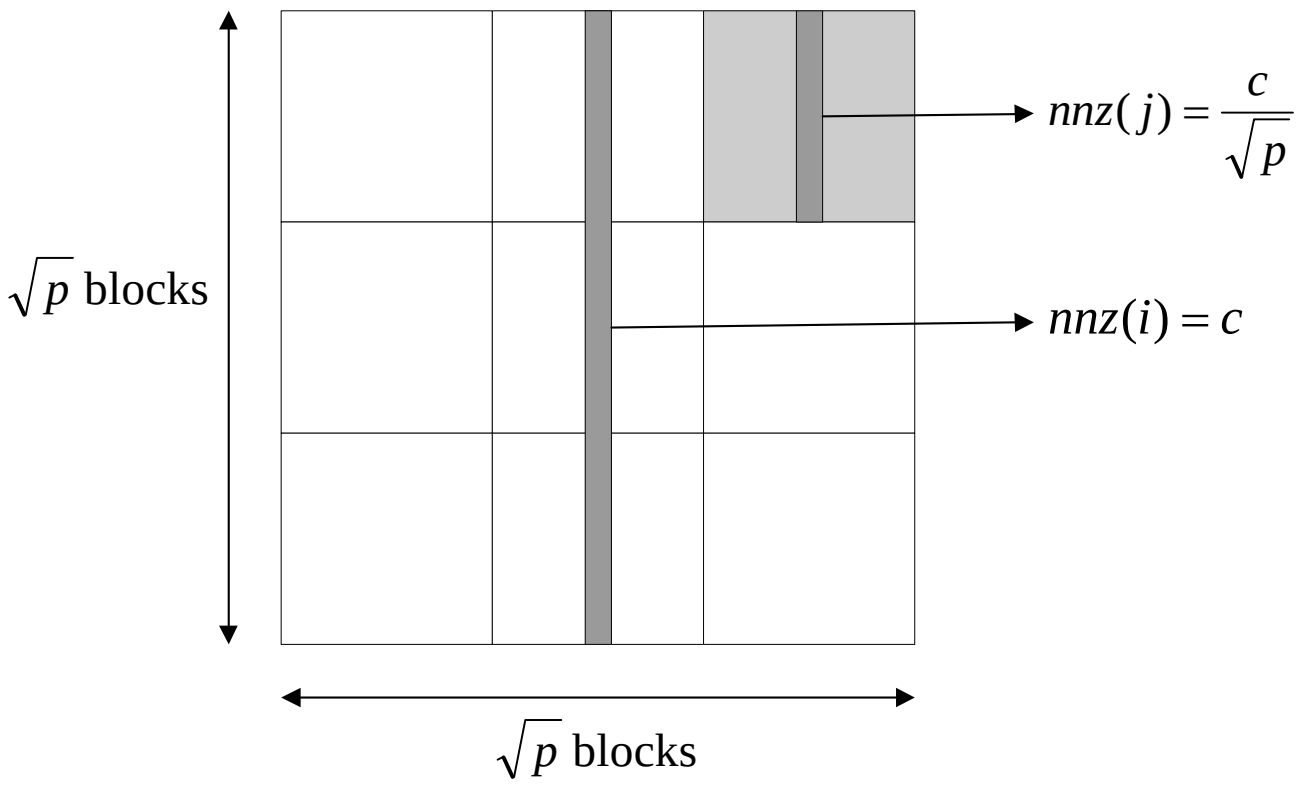

Figure 4: 2D Sparse Matrix Decomposition

| | | | | | | | | | | | |
|---|---|---|---|---|---|---|---|---|---|---|---|
| JC | = | 1 | 3 | 3 | 3 | 3 | 3 | 3 | 4 | 5 | 5 |
| | | ↓ | | | | | | ↓ | ↓ | | |
| IR | = | 6 | 8 | | | | | 4 | 2 | | |
| NUM | = | 0.1 | 0.2 | | | | | 0.3 | 0.4 | | |

Figure 5: Matrix **A** in CSC format

Figure 4 shows that the average number of nonzeros in a single column of a submatrix, $nnz(j)$, goes to zero as $p$ increases. Storing a graph using CSC is similar to using adjacency lists. The column-pointers array represents the vertices, and the row-indices array represents their adjacencies. In that sense, CSC is a vertex based data structure, making it suitable for 1D (vertex) partitioning of the graph. 2D partitioning, on the other hand, is based on edges. Therefore, using CSC with 2D distributed data is forcing a vertex based representation on edge distributed data. The result is unnecessary replication of column pointers (vertices) on each processor along the processor column.

The inefficiency of CSC leads to a more fundamental problem: any algorithm that uses CSC and scans all the columns is not scalable for hypersparse matrices. Even without any communication at all, such an algorithm cannot scale for $n\sqrt{p} \geq max\{\text{flops}, nnz\}$. Sparse matrix-vector and sparse matrix-matrix multiplication algorithms scan column indices. For these operations, any data structure that depends on the matrix dimension (such as CSR or CSC) is asymptotically too wasteful for submatrices.

### *2.3. DCSC Data Structure*

We use a new data structure for our sequential hypersparse matrix-matrix multiplication. This structure, called ***DCSC*** for doubly compressed sparse columns, has the

| **A**.I | **A**.J | **A**.V |
|---|---|---|
| 6 | 1 | 0.1 |
| 8 | 1 | 0.2 |
| 4 | 7 | 0.3 |
| 2 | 8 | 0.4 |

Figure 6: Matrix **A** in Triples format

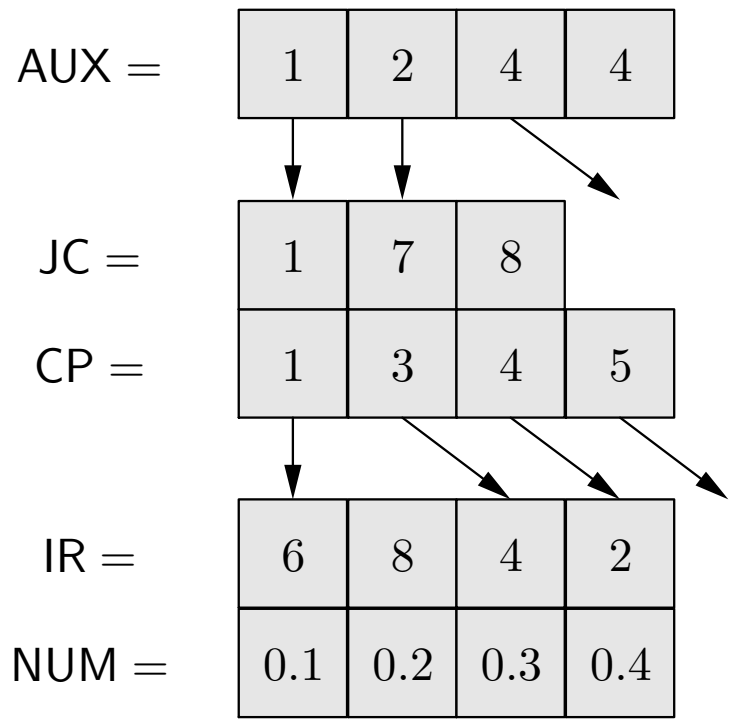

Figure 7: Matrix **A** in DCSC format

following properties:

1. It uses $O(nnz)$ storage.
2. It lets the hypersparse algorithm scale with increasing sparsity.
3. It supports fast access to columns of the matrix, when necessary.

For an example, consider the 9-by-9 matrix with 4 non-zeros whose triples representation is given in Figure 6. Figure 5 showns its CSC storage, which includes repetitions and redundancies in the column pointers array (JC). Our new data structure compresses the JC array to avoid repetitions, giving the CP(column pointers) array of DCSC as shown in Figure 7. DCSC is essentially a sparse array of sparse columns, whereas CSC is a dense array of sparse columns.

After removing repetitions, CP[$i$] does no longer refer to the $i$th column. A new JC array, which is parallel to CP, gives us the column numbers. Although our H - _GEMM algorithm does not need column indexing, DCSC supports fast column indexing for completeness. Whenever column indexing is needed, we construct an AUX array that contains pointers to nonzero columns (columns that have at least one nonzero element). Each entry in AUX refers to a $\lceil n/nzc \rceil$-sized chunk of columns, pointing to the first nonzero column in that chunk (there might be none). The storage requirement of DCSC is $O(nnz)$ since $|\text{NUM}| = |\text{IR}| = nnz$, $|\text{JC}| = nzc$, $|\text{CP}| = nzc + 1$, and $|\text{AUX}| \approx nzc$.

In our implementation, the AUX array is a temporary work array that is contructed on demand, only when an operation requires repetitive use of it. This keeps the storage and copying costs low. The time to construct AUX is only $O(nzc)$, which is subsumed by the cost of multiplication.

### *2.4. A Sequential Algorithm to Multiply Hypersparse Matrices*

The sequential hypersparse algorithm (H _GEMM) is based on outer product multiplication. Therefore, it requires fast access to rows of matrix **B**. This could be accomplished by having each input matrix represented in DCSC and also in DCSR

$$\mathbf{A} = \begin{array}{c} \\ 1 \\ 2 \\ 3 \\ 4 \\ 5 \\ 6 \end{array} \begin{array}{c} \begin{array}{cccccc} 1 & 2 & 3 & 4 & 5 & 6 \end{array} \\ \begin{pmatrix} \times & & & \times & & \\ \times & \times & & & & \\ & & \times & \times & & \times \\ \times & & & \times & & \\ & \times & & & & \\ & & \times & & & \end{pmatrix} \end{array}, \quad \mathbf{B} = \begin{array}{c} \\ 1 \\ 2 \\ 3 \\ 4 \\ 5 \\ 6 \end{array} \begin{array}{c} \begin{array}{cccccc} 1 & 2 & 3 & 4 & 5 & 6 \end{array} \\ \begin{pmatrix} \times & & & \times & & \\ & & & & & \\ & & \times & & \times & \\ \times & & & \times & & \\ & \times & & & \times & \times \\ & & \times & & \times & \times \end{pmatrix} \end{array}$$

Figure 8: Nonzero structures of operands **A** and **B**

(doubly compressed sparse rows), which is the same as the transpose in DCSC. This method, which we described in an early version of this work [26], doubles the storage but does not change the asymptotic space and time complexities. Here, we describe a more practical version where **B** is transposed as a preprocessing step, at a cost of trans(**B**). The actual cost of transposition is either $O(n+nnz(\mathbf{B}))$ or $O(nnz(\mathbf{B}) \lg nnz(\mathbf{B}))$, depending on the implementation.

The idea behind the H _GEMM algorithm is to use the outer product formulation of matrix multiplication efficiently. The first observation about DCSC is that the JC array is already sorted. Therefore, **A**.JC is the sorted indices of the columns that contain at least one nonzero and similarly $\mathbf{B^T}$.JC is the sorted indices of the rows that contain at least one nonzero. In this formulation, the *i*th column of **A** and the *i*th row of **B** are multiplied to form a rank-1 matrix. The naive algorithm does the same procedure for all values of *i* and gets *n* different rank-1 matrices, adding them to the resulting matrix **C** as they become available. Our algorithm has a preprocessing step that finds intersection $\text{Isect} = \mathbf{A}.\text{JC} \cap \mathbf{B^T}.\text{JC}$, which is the set of indices that participate nontrivially in the outer product.

The preprocessing takes $O(nzc(\mathbf{A}) + nzr(\mathbf{B}))$ time as $|\mathbf{A}.\text{JC}| = nzc(\mathbf{A})$ and $|\mathbf{B^T}.\text{JC}| = nzr(\mathbf{B})$. The next phase of our algorithm performs $|\text{Isect}|$ cartesian products, each of which generates a fictitious list of size $nnz(\mathbf{A}(:,i)) \cdot nnz(\mathbf{B}(i,:))$. The lists can be generated sorted, because all the elements within a given column are sorted according to their row indices (i.e. $\text{IR}(\text{JC}(i))...\text{IR}(\text{JC}(i)+1)$ is a sorted range). The algorithm merges those sorted lists, summing up the intermediate entries having the same $(row_id, col_id)$ index pair, to form the resulting matrix **C**. Therefore, the second phase of H - _GEMM is similar to multiway merging [27]. The only difference is that we never explicitly construct the lists; we compute their elements one-by-one on demand.

Figure 9 shows the setup for the matrices from Figure 8. As $\mathbf{A}.\text{JC} = \{1, 2, 3, 4, 6\}$ and $\mathbf{B^T}.\text{JC} = \{1, 3, 4, 5, 6\}$, $\text{Isect} = \{1, 3, 4, 6\}$ for this product. The algorithm does not touch the shaded elements, since they do not contribute to the output.

The merge uses a priority queue (represented as a heap) of size *ni*, which is the size of Isect, the number of indices *i* for which $\mathbf{A}(:,i) \neq \emptyset$ and $\mathbf{B}(i,:) \neq \emptyset$. The value in a heap entry is its NUM value and the key is a pair of indices $(i, j)$ in column-major order. The idea is to repeatedly extract the entry with minimum key from the heap and insert another element from the list that the extracted element originally came from. If there are multiple elements in the lists with the same key, then their values are added on the fly. If we were to explicitly create *ni* lists instead of doing the computation on the fly, we would get the lists shown in the right side of Figure 9, which are sorted from bottom to top. For further details of multiway merging, consult Knuth [27].

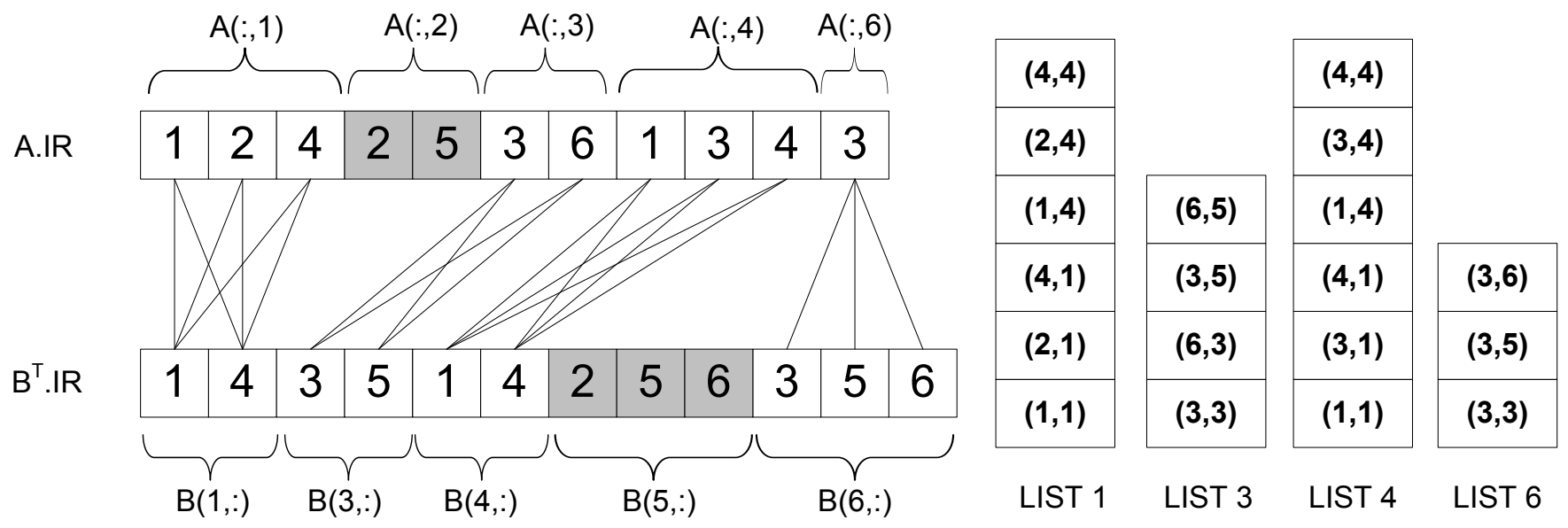

Figure 9: Cartesian product and the multiway merging analogy

The time complexity of this phase is $O(\text{flops} \cdot \lg ni)$, and the space complexity is $O(nnz(\mathbf{C}) + ni)$. The output is a stack of NUM values in column-major order. The $nnz(\mathbf{C})$ term in the space complexity comes from the output, and the flops term in the time complexity comes from the observation that

$$\sum_{i \in \text{Isect}} nnz(\mathbf{A}(:, i)) \cdot nnz(\mathbf{B}(i, :)) = \text{flops}.$$

The final phase of the algorithm constructs the DCSC structure from this column-major-ordered stack. This requires $O(nnz(\mathbf{C}))$ time and space.

The overall time complexity of our algorithm is $O(nzc(\mathbf{A}) + nzr(\mathbf{B}) + \text{flops} \cdot \lg ni)$, plus the preprocessing time to transpose matrix **B**. Note that $nnz(\mathbf{C})$ does not appear in this bound, since $nnz(\mathbf{C}) \leq \text{flops}$. We opt to keep the cost of transposition separate, because our parallel 2D block SpGEMM will amortize this transposition of each block over $\sqrt{p}$ uses of that block. Therefore, the cost of transposition will be negligible in practice. The space complexity is $O(nnz(\mathbf{A}) + nnz(\mathbf{B}) + nnz(\mathbf{C}))$. The time complexity does not depend on $n$, and the space complexity does not depend on flops.

Figure 10 gives the pseudocode for the whole algorithm. It uses two subprocedures: C M -I generates the next element from the $i$th fictitious list and inserts it to the heap PQ, and I -L increments the pointers of the $i$th fictitious list or deletes the list from the intersection set if it is empty.

To justify the extra logarithmic factor in the flops term, we briefly analyze the complexity of each submatrix multiplication in the parallel 2D block SpGEMM. Our parallel 2D block SpGEMM performs $p\sqrt{p}$ submatrix multiplications, since each submatrix of the output is computed using $\mathbf{C}_{ij} = \sum_{k=1}^{\sqrt{p}} \mathbf{A}_{ik}\mathbf{B}_{kj}$. Therefore, with increasing number of processors and under perfect load balance, flops scale with $1/p\sqrt{p}$, *nnz* scale with $1/p$, and $n$ scales with $1/\sqrt{p}$. Figure 11 shows the trends of these three complexity measures as $p$ increases. The graph shows that the $n$ term becomes the bottleneck after around 50 processors and flops becomes the lower-order term. In contrast to the

```
C : ℝ^{S(m×n)} = H _GEMM(A : ℝ^{S(n×k)}, B^T : ℝ^{S(n×k)})
Isect ← I (A.JC, B^T.JC)
for j ← 1 to |Isect|
    do C M -I (A, B^T, PQ, Isect, j)
       I -L (Isect, j)
while I N F (Isect)
    do (key, value) ← E -M (PQ)
       (product, i) ← U P (value)
       if key ≠ T (Q)
          then E (Q, key, product)
          else U T (Q, product)
       if I N E (Isect(i))
          then C M -I (A, B^T, PQ, lists, Isect, i)
               I -L (Isect, i)
C -D (Q)
```

Figure 10: Pseudocode for hypersparse matrix-matrix multiplication algorithm

classical algorithm, our H _GEMM algorithm becomes independent of $n$, by putting the burden on the flops instead.

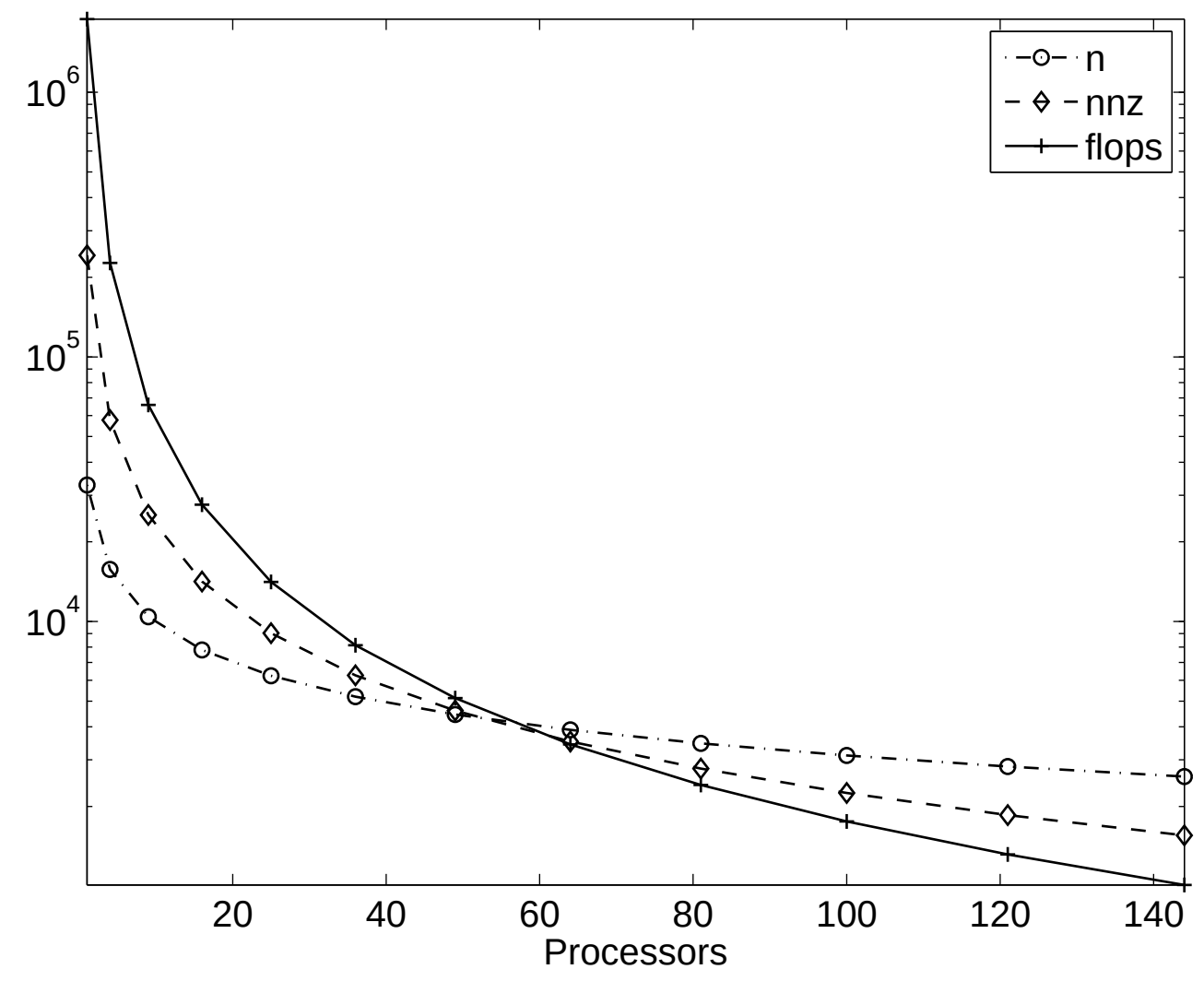

Figure 11: Trends of different complexity measures for submatrix multiplications as $p$ increases. The inputs are randomly permuted RMAT matrices (scale 15 with an average of 8 nonzeros per column) that are successively divided into $(n/\sqrt{p}) \times (n/\sqrt{p})$. The counts are averaged over all submatrix multiplications.

## 3. Parallel Algorithms for Sparse GEMM

This section describes parallel algorithms for multiplying two sparse matrices in parallel on $p$ processors, which we call PSpGEMM. The design of our algorithms is motivated by distributed memory systems, but expect them to perform well in shared memory too, as they avoid hot spots and load imbalances by ensuring proper work distribution among processors. Like most message passing algorithms, they can be implemented in the partitioned global address space (PGAS) model as well.

### *3.1. 1D Decomposition*

We assume the data is distributed to processors in block rows, where each processor receives $m/p$ consecutive rows. We write $\mathbf{A}_i = \mathbf{A}(ip : (i+1)p-1, :)$ to denote the block row owned by the $i$th processor. To simplify the algorithm description, we use $\mathbf{A}_{ij}$ to denote $\mathbf{A}_i(:, jp : (j+1)p-1)$, the $j$th block column of $\mathbf{A}_i$, although block rows are not physically partitioned.

$$\mathbf{A} = \left(\begin{array}{c} \mathbf{A}_1 \\ \hline \vdots \\ \hline \mathbf{A}_p \end{array}\right) = \left(\begin{array}{c|c|c} \mathbf{A}_{11} & \dots & \mathbf{A}_{1p} \\ \hline \vdots & \ddots & \vdots \\ \hline \mathbf{A}_{p1} & \dots & \mathbf{A}_{pp} \end{array}\right), \mathbf{B} = \left(\begin{array}{c} \mathbf{B}_1 \\ \hline \vdots \\ \hline \mathbf{B}_p \end{array}\right) \tag{3}$$

For each processor $P(i)$, the computation is:

$$\mathbf{C}_i = \mathbf{C}_i + \mathbf{A}_i \, \mathbf{B} = \mathbf{C}_i = \mathbf{C}_i + \sum_{j=1}^{p} \mathbf{A}_{ij} \, \mathbf{B}_j$$

### *3.2. 2D Decomposition*

Our 2D parallel algorithms, Sparse Cannon and Sparse SUMMA, use the hypersparse algorithm, which has complexity $O(nzc(\mathbf{A}) + nzr(\mathbf{B}) + \text{flops} \cdot \lg ni)$, as shown in Section 2.2, for multiplying submatrices. Processors are logically organized on a square $\sqrt{p} \times \sqrt{p}$ mesh, indexed by their row and column indices so that the $(i, j)$th processor is denoted by $P(i, j)$. Matrices are assigned to processors according to a 2D block decomposition. Each node gets a submatrix of dimensions $(n/\sqrt{p}) \times (n/\sqrt{p})$ in its local memory. For example, $\mathbf{A}$ is partitioned as shown below and $\mathbf{A}_{ij}$ is assigned to processor $P(i, j)$.

$$\mathbf{A} = \left(\begin{array}{c|c|c} \mathbf{A}_{11} & \dots & \mathbf{A}_{1\sqrt{p}} \\ \hline \vdots & \ddots & \vdots \\ \hline \mathbf{A}_{\sqrt{p}1} & \dots & \mathbf{A}_{\sqrt{p}\sqrt{p}} \end{array}\right) \tag{4}$$

For each processor $P(i)$, the computation is:

$$\mathbf{C}_{ij} = \sum_{k=1}^{\sqrt{p}} \mathbf{A}_{ik} \, \mathbf{B}_{kj}$$

$\mathbf{C} : \mathbb{R}^{P(S(n) \times n)}$ = B 1D_PS GEMM($\mathbf{A} : \mathbb{R}^{P(S(n) \times n)}, \mathbf{B} : \mathbb{R}^{P(S(n) \times n)}$)

**for** all processors $P(i)$ in parallel
    **do** I (SPA)
        **for** $j \leftarrow 1$ **to** $p$
            **do** B ($\mathbf{B}_j$)
                **for** $k \leftarrow 1$ **to** $n/p$
                    **do** L (SPA, $\mathbf{C}_i(k, :)$)
                        SPA $\leftarrow$ SPA + $\mathbf{A}_{ij}(k, :)\,\mathbf{B}_j$
                        U (SPA, $\mathbf{C}_i(k, :)$)

Figure 12: Operation $\mathbf{C} \leftarrow \mathbf{AB}$ using block row Sparse 1D algorithm

L -C -S ($\mathbf{Local} : \mathbb{R}^{S(n \times n)}, s$)

S ($\mathbf{Local}, P(i, (j - s) \bmod \sqrt{p})$) ▷ This is processor $P(i, j)$
R ($\mathbf{Temp}, P(i, (j + s) \bmod \sqrt{p})$)
**Local** $\leftarrow$ **Temp**

Figure 13: Circularly shift left by $s$ along the processor row

### *3.3. Sparse 1D Algorithm*

The row-wise SpGEMM forms one row of **C** at a time, and each processor may potentially need to access all of **B** to form a single row of **C**. However, only a portion of **B** is locally available at any time in parallel algorithms. The algorithm, thus, performs multiple iterations to fully form one row of **C**. For accumulating the nonzeros of the current active row of **C**, the algorithm uses a special structure called the sparse accumulator (SPA) [22] that performs accumulation in linear time. Figure 12 shows the pseudocode of the algorithm.

### *3.4. Sparse Cannon*

Our first 2D algorithm is based on Cannon's algorithm for dense matrices [28]. The pseudocode of the algorithm is given in Figure 15. Sparse Cannon, although elegant, is not our choice of algorithm for the final implementation, as it is hard to generalize to non-square grids, non-square matrices, and matrices whose dimensions are not perfectly divisible by grid dimensions.

### *3.5. Sparse SUMMA*

SUMMA [29] is a memory efficient, easy to generalize algorithm for parallel dense matrix multiplication. It is the algorithm used in parallel BLAS [30]. As opposed to Cannon's algorithm, it allows a tradeoff to be made between latency cost and memory by varying the degree of blocking. The algorithm, illustrated in Figure 16, proceeds

U -C -S (**Local** : $\mathbb{R}^{S(n\times n)}$, $s$)

S (**Local**, $P((i-s) \bmod \sqrt{p}, j)$) ▷ This is processor $P(i, j)$
R (**Temp**, $P((i+s) \bmod \sqrt{p}, j)$)
**Local** ← **Temp**

Figure 14: Circularly shift up by $s$ along the processor column

**C** : $\mathbb{R}^{P(S(n\times n))}$ = C _PS GEMM(**A** : $\mathbb{R}^{P(S(n\times n))}$, **B** : $\mathbb{R}^{P(S(n\times n))}$)

**for** all processors $P(i, j)$ in parallel
  **do** L -C -S ($\mathbf{A}_{ij}$, $i-1$)
    U -C -S ($\mathbf{B}_{ij}$, $j-1$)
**for** all processors $P(i, j)$ in parallel
  **do for** $k \leftarrow 1$ **to** $\sqrt{p}$
    **do** $\mathbf{C}_{ij} \leftarrow \mathbf{C}_{ij} + \mathbf{A}_{ij}\,\mathbf{B}_{ij}$
      L -C -S ($\mathbf{A}_{ij}$, 1)
      U -C -S ($\mathbf{B}_{ij}$, 1)

Figure 15: Operation **C** ← **AB** using Sparse Cannon

in $k/b$ stages. At each stage, $\sqrt{p}$ active row processors broadcast $b$ columns of **A** simultaneously along their rows and $\sqrt{p}$ active column processors broadcast $b$ rows of **B** simultaneously along their columns.

Sparse SUMMA, which is a sparse generalization of SUMMA, is our algorithm of choice for our final implementation, because it is easy to generalize to non-square matrices, and to matrices whose dimensions are not perfectly divisible by grid dimensions.

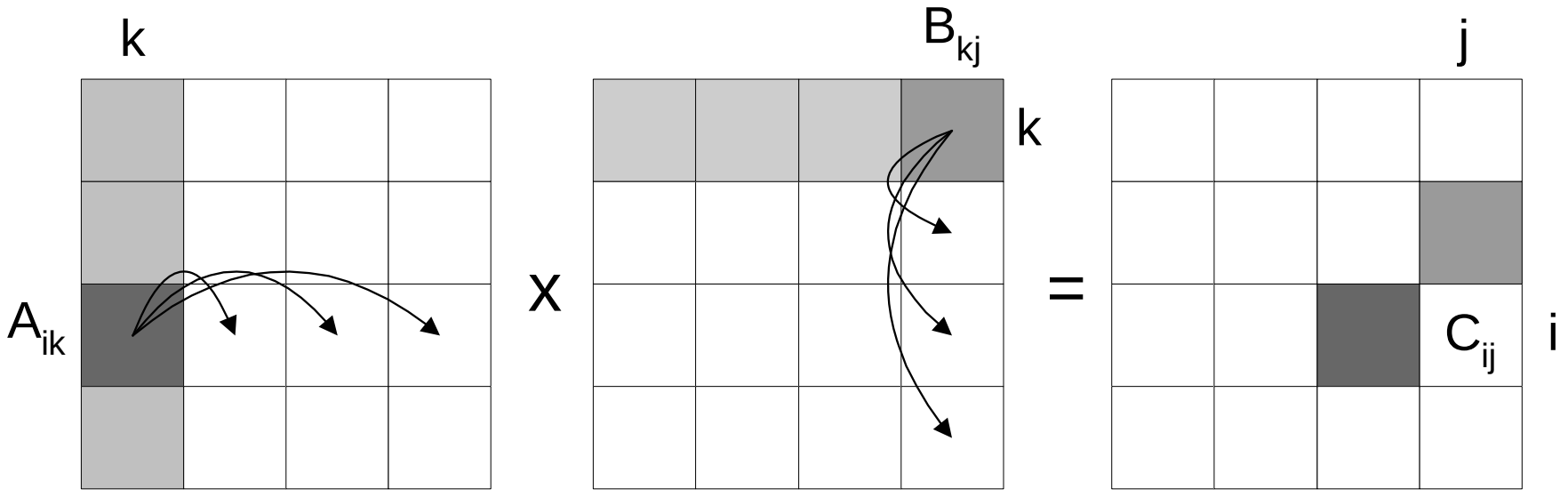

Figure 16: Sparse SUMMA Execution ($b = n/\sqrt{p}$)

## 4. Analysis of Parallel Algorithms

In this section, we analyze the parallel performance of our algorithms, and show that they scale better than existing 1D algorithms in theory. We begin by introducing our parameters and model of computation. Then, we present a theoretical analysis showing that 1D decomposition, at least with the current algorithm, is not sufficient for PSpGEMM to scale. Finally, we analyze our 2D algorithms in depth.

In our analysis, the cost of one floating-point operation, along with the cost of cache misses and memory indirections associated with the operation, is denoted by $\gamma$, measured in nanoseconds. The latency of sending a message over the communication interconnect is $\alpha$, and the inverse bandwidth is $\beta$, measured in nanoseconds and nanoseconds per word transfered, respectively. The running time of a parallel algorithm on $p$ processors is given by

$$T_p = T_{comm} + T_{comp},$$

where $T_{comm}$ denotes the time spent in communication and $T_{comp}$ is the time spent during local computation phases. $T_{comm}$ includes both the latency (delay) costs and the actual time it takes to transfer the data words over the network. Hence, the cost of transmitting $h$ data words in a communication phase is

$$T_{comm} = \alpha + h\beta.$$

The sequential work of SpGEMM, unlike dense GEMM, depends on many parameters. This makes parallel scalability analysis a tough process. Therefore, we restrict our analysis to sparse matrices following the Erdős-Rényi graph model explained in Section 5.1. Consequently, the analysis is probabilistic, exploiting the independent and identical distribution of nonzeros. When we talk about quantities such as nonzeros per subcolumn, we mean the expected number of nonzeros. Our analysis assumes that there are $c > 0$ nonzeros per row/column. The sparsity parameter $c$, albeit oversimplifying, is useful for analysis purposes, since it makes different parameters comparable to each other. For example, if $\mathbf{A}$ and $\mathbf{B}$ both have sparsity $c$, then $nnz(\mathbf{A}) = cn$ and $\text{flops}(\mathbf{AB}) = c^2 n$. It also allows us to decouple the effects of load imbalances from the algorithm analysis because the nonzeros are assumed to be evenly distributed across processors.

The lower bound on sequential SpGEMM is $\Omega(\text{flops}) = \Omega(c^2 n)$. This bound is achieved by some row-wise and column-wise implementations [21, 22], provided that $c \geq 1$. Gustavson's classical algorithm implemented using CSR is the natural kernel to be used in the 1D algorithm where data is distributed by rows. As mentioned earlier, it has an asymptotic complexity of

$$O(n + nnz(\mathbf{A}) + \text{flops}) = O(n + cn + c^2 n) = \Theta(c^2 n).$$

Therefore, we take the sequential work ($W$) to be $\gamma c^2 n$ in our analysis.

### *4.1. Scalability of the 1D Algorithm*

We begin with a theoretical analysis whose conclusion is that 1D decomposition is not sufficient for PSpGEMM to scale. In B 1D_PS GEMM, each processor sends and receives $p - 1$ point-to-point messages of size $nnz(\mathbf{B})/p$. Therefore,

$$T_{comm} = (p-1)\,(\alpha + \beta\,\frac{nnz(\mathbf{B})}{p}) = \Theta(p\,\alpha + \beta\,c\,n). \tag{5}$$

We previously showed that the B 1D_PS GEMM algorithm is unscalable with respect to both communication and computation costs [31]. In fact, the computational cost per processor is constant with increasing number of processors, disabling any speedup. This is because the cost of SPA loading and unloading, which is not amortized by the number of nonzero arithmetic operations in general, dominate the computational time. The current S -P implementation [7] by-passes this problem by all-to-all broadcasting nonzeros of the **B** matrix, so that the whole **B** matrix is essentially assembled at each processor. This avoids the cost of loading and unloading SPA at every stage, but it uses $nnz(\mathbf{B})$ memory at each processor.

### *4.2. Scalability of the 2D Algorithms*

In this section, we provide an in-depth theoretical analysis of our parallel 2D SpGEMM algorithms, and conclude that they scale significantly better than their 1D counterparts. Although our analysis is limited to the Erdős-Rényi model, its conclusions are strong enough to be convincing.

In C _PS GEMM, each processor sends and receives $\sqrt{p}-1$ point-to-point messages of size $nnz(\mathbf{A})/p$, and $\sqrt{p}-1$ messages of size $nnz(\mathbf{B})/p$. Therefore, the communication cost per processor is

$$T_{comm} = \sqrt{p}\,\Big(2\,\alpha + \beta\,(\frac{nnz(\mathbf{A}) + nnz(\mathbf{B})}{p})\Big) = \Theta(\alpha\,\sqrt{p} + \frac{\beta\,c\,n}{\sqrt{p}}). \tag{6}$$

The average number of nonzeros in a column of a local submatrix $\mathbf{A}_{ij}$ is $c/\sqrt{p}$. Therefore, for a submatrix multiplication $\mathbf{A}_{ik}\mathbf{B}_{kj}$,

$$ni(\mathbf{A}_{ik}, \mathbf{B}_{kj}) = \min\{1, \frac{c^2}{p}\}\frac{n}{\sqrt{p}} = \min\{\frac{n}{\sqrt{p}}, \frac{c^2\,n}{p\,\sqrt{p}}\},$$

$$\text{flops}(\mathbf{A}_{ik}\mathbf{B}_{kj}) = \frac{\text{flops}(\mathbf{AB})}{p\,\sqrt{p}} = \frac{c^2\,n}{p\,\sqrt{p}},$$

$$T_{mult} = \sqrt{p}\,\Big(2\,\min\{1, \frac{c}{\sqrt{p}}\}\frac{n}{\sqrt{p}} + \frac{c^2 n}{p\,\sqrt{p}}\,\lg\Big(\min\{\frac{n}{\sqrt{p}}, \frac{c^2 n}{p\,\sqrt{p}}\}\Big)\Big).$$

The probability of a single column of $\mathbf{A}_{ik}$ (or a single row of $\mathbf{B}_{kj}$) having at least one nonzero is $\min\{1, c/\sqrt{p}\}$ where 1 covers the case $p \le c^2$ and $c/\sqrt{p}$ covers the case $p > c^2$.

The overall cost of additions, using $p$ processors, and Brown and Tarjan's $O(m\lg n/m)$ algorithm [32] for merging two sorted lists of size $m$ and $n$ (for $m < n$), is

$$T_{add} = \sum_{i=1}^{\sqrt{p}}\Big(\frac{\text{flops}}{p\,\sqrt{p}}\,\lg i\Big) = \frac{\text{flops}}{p\,\sqrt{p}}\,\lg\prod_{i=1}^{\sqrt{p}} i = \frac{\text{flops}}{p\,\sqrt{p}}\,\lg(\sqrt{p}\,!).$$

Note that we might be slightly overestimating, since we assume flops/ $nnz(\mathbf{C}) \approx 1$ for simplicity. From Stirling's approximation and asymptotic analysis, we know that $\lg(n\,!) = \Theta(n \lg n)$ [33]. Thus, we get:

$$T_{add} = \Theta\Big(\frac{\text{flops}}{p\sqrt{p}}\ \sqrt{p}\ \lg\ \sqrt{p}\Big) = \Theta\Big(\frac{c^2 n \lg\ \sqrt{p}}{p}\Big).$$

There are two cases to analyze: $p > c^2$ and $p \leq c^2$. Since scalability analysis is concerned with the asymptotic behavior as $p$ increases, we just provide results for the $p > c^2$ case. The total computation cost $T_{comp} = T_{mult} + T_{add}$ is

$$T_{comp} = \gamma\Big(\frac{c\,n}{\sqrt{p}} + \frac{c^2 n}{p}\lg(\frac{c^2 n}{p\sqrt{p}}) + \frac{c^2 n \lg\ \sqrt{p}}{p}\Big) = \gamma\Big(\frac{c\,n}{\sqrt{p}} + \frac{c^2 n}{p}\lg(\frac{c^2 n}{p})\Big). \tag{7}$$

In this case, parallel efficiency is

$$E = \frac{W}{p\,(T_{comp} + T_{comm})} = \frac{\gamma c^2 n}{(\gamma + \beta)\,c\,n\ \sqrt{p} + \gamma c^2 n\ \lg(\frac{c^2 n}{p}) + \alpha\,p\ \sqrt{p}}. \tag{8}$$

Scalability is not perfect and efficiency deteriorates as $p$ increases due to the first term. Speedup is, however, not bounded, as opposed to the 1D case. In particular, $\lg(c^2 n/p)$ becomes negligible as $p$ increases and scalability due to latency is achieved when $\gamma c^2 n \propto \alpha\,p\sqrt{p}$, where it is sufficient for $n$ to grow on the order of $p^{1.5}$. The biggest bottleneck for scalability is the first term in the denominator, which scales with $\sqrt{p}$. Consequently, two different scaling regimes are likely to be present: A close to linear scaling regime until the first term starts to dominate the denominator and a $\sqrt{p}$-scaling regime afterwards.

Compared to the 1D algorithms, Sparse Cannon both lower the degree of unscalability due to bandwidth costs and mitigate the bottleneck of computation. This makes overlapping communication with computation more promising.

Sparse SUMMA, like dense SUMMA, incurs an extra cost over Cannon for using row-wise and col-wise broadcasts instead of nearest-neighbor communication, which might be modeled as an additional $O(\lg p)$ factor in communication cost. Other than that, the analysis is similar to sparse Cannon and we omit the details. Using the DCSC data structure, the expected cost of fetching $b$ consecutive columns of a matrix $\mathbf{A}$ is $b$ plus the size (number of nonzeros) of the output [26]. Therefore, the algorithm asymptotically has the same computation cost for all values of $b$.

## 5. Performance Modeling of Parallel Algorithms

In this section, we first project the estimated speedup of 1D and 2D algorithms in order to evaluate their prospects in practice. We use a quasi-analytical performance model where we first obtain realistic values for the parameters $(\gamma, \beta, \alpha)$ of the algorithm performance, then use them in our projections. In the second part, we perform another modeling study where we simulate the execution of Sparse SUMMA using an actual implementation of the H _GEMM algorithm. This modeling study concludes that H _GEMM is scalable with increasing hypersparsity, suggesting that it is a suitable algorithm to be the sequential kernel of a 2D parallel SpGEMM.

### *5.1. Sparse Matrix Models*

For testing and analysis, we have extensively used three main models: the R-MAT model, the Erdős-Rényi random graph model, and the regular 3D grid model.

#### *5.1.1. Synthetic R-MAT Graphs*

The R-MAT matrices represent the adjacency structure of scale-free graphs, generated using repeated Knonecker products [34, 35]. R-MAT models the behavior of several real-world graphs such as the WWW graph, small world graphs, and citation graphs. We have used an implementation based on Kepner's vectorized code [36], which generates directed graphs. Unless otherwise stated, R-MAT matrices used in our experiments have an average of degree of 8, meaning that there will be approximately $8\,n$ nonzeros in the adjacency matrix. The parameters for the generator matrix are $a = 0.6$, and $b = c = d = 0.13$. As the generator matrix is 2-by-2, R-MAT matrices have dimensions that are powers of two. An R-MAT graph with ***scale*** $l$ has $n = 2^l$ vertices.

#### *5.1.2. Erdős-Rényi Random Graphs*

An Erdős-Rényi random graph $G(n, p)$ has $n$ vertices, each of the possible $n^2$ edges in the graph exists with fixed probability $p$, independent of the other edges [37]. In other words, each edge has an equally likely chance to exist. A matrix modeling the Erdős-Rényi graph $G(n, p)$ is expected to have with $n^2/p$ nonzeros, independently and identically distributed (i.i.d.) across the matrix. Erdős-Rényi random graphs can be generated using the `sprand` function of M .

#### *5.1.3. Regular 3D Grids*

As the representative of regular grid graphs, we have used matrices arising from graphs representing the 3D 7-point finite difference mesh (`grid3d`). These input matrices, which are generated using the M Mesh Partitioning and Graph Separator Toolbox [38], are highly structured block diagonal matrices.

### *5.2. Estimated Speedup of Parallel Algorithms*

This study estimates the speedup of 1D and 2D algorithms by using a quasi-analytic model that projects the performance on large systems using realistic values for the performance parameters.

In order to obtain a realistic value for $\gamma$, we performed multiple runs on an AMD Opteron 8214 (Santa Rosa) processor using matrices of various dimensions and sparsity; estimating the constants using non-linear regression. One surprising result is the order of magnitude difference in the constants between sequential kernels. The classical algorithm, which is used as the 1D SpGEMM kernel, has $\gamma = 293.6$ nsec, whereas H _GEMM, which is used as the 2D kernel, has $\gamma = 19.2$ nsec. We attribute the difference to cache friendliness of the hypersparse algorithm. The interconnect supports $1/\beta = 1$ GB/sec point-to-point bandwidth, and a maximum of $\alpha = 2.3$ microseconds latency, both of which are achievable on TACC's Ranger Cluster. The communication parameters ignore network contention.

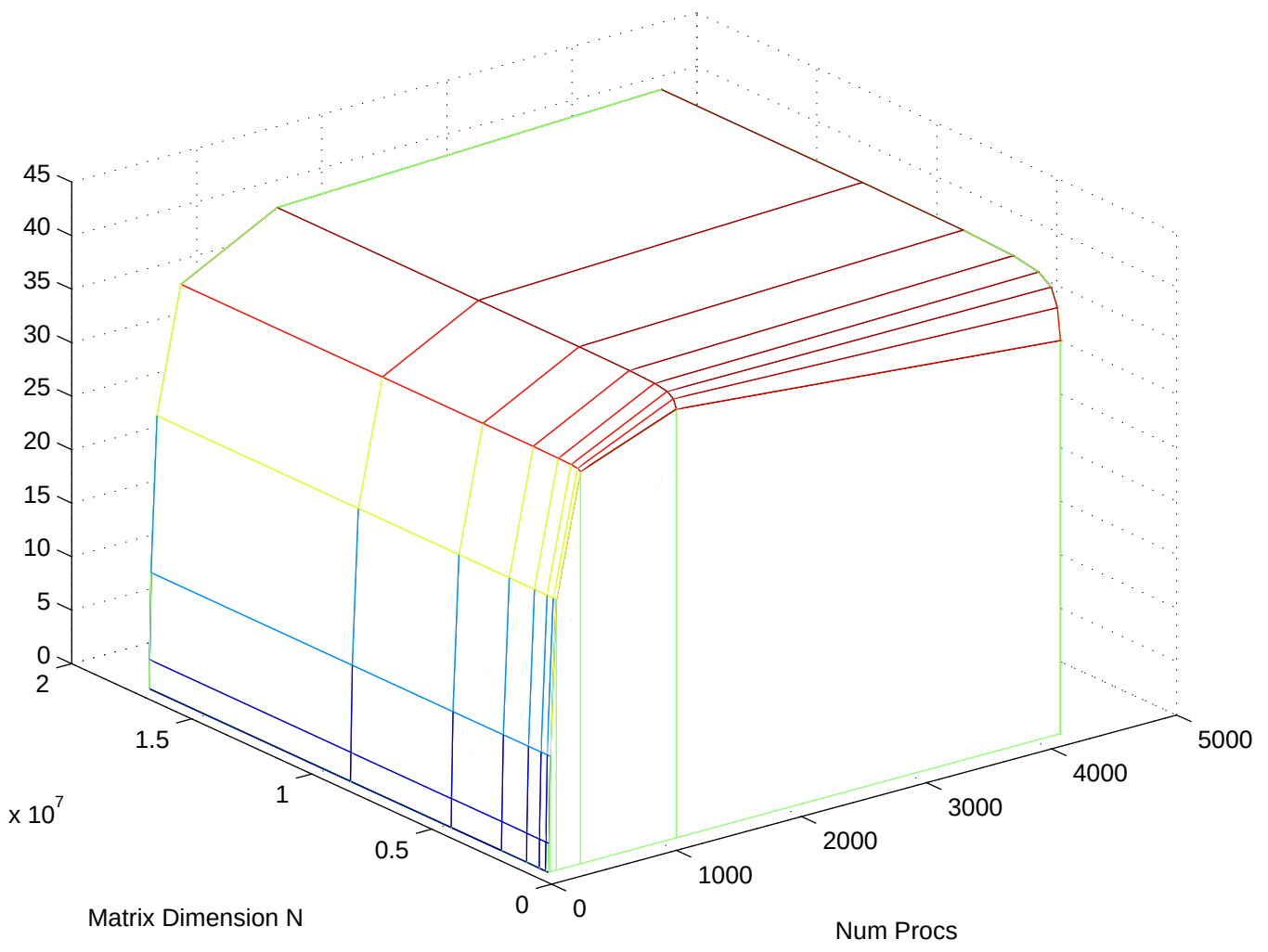

Figure 17: Modeled speedup of Synchronous Sparse 1D algorithm

Figures 17 and 18 show the modeled speedup of B 1D_PS GEMM and C - _PS GEMM for matrix dimensions from $n = 2^{17}$ to $2^{24}$ and number of processors from $p = 1$ to 4096. The inputs are Erdős-Rényi graphs.

We see that B 1D_PS GEMM's speedup does not go beyond 50x, even on larger matrices. For relatively small matrices, having dimensions $n = 2^{17} - 2^{20}$, it starts slowing down after a thousand processors, where it achieves less than 40x speedup. On the other hand, C _PS GEMM shows increasing and almost linear speedup for up to 4096 processors, even though the slope of the curve is less than one. It is crucial to note that the projections for the 1D algorithm are based on the memory inefficient implementation that performs an all-to-all broadcast of **B**. This is because the original memory efficient algorithm given in Section 3.1 actually slows down as $p$ increases.

It is worth explaining one peculiarity. The modeled speedup turns out to be higher for smaller matrices than for bigger matrices. Remember that communication requirements are on the same order as computational requirements for parallel SpGEMM. Intuitively, the speedup should be independent of the matrix dimension in the absence of load imbalance and network contention, but since we are estimating the speedup with respect to the optimal sequential algorithm, the overheads associated with the hypersparse algorithm are bigger for larger matrices. The bigger the matrix dimension, the slower the hypersparse algorithm is with respect to the optimal algorithm, due to the extra logarithmic factor. Therefore, speedup is better for smaller matrices in theory. This is not the case in practice, because the peak bandwidth is usually not achieved for small sized data transfers and load imbalances are severer for smaller matrices. Section 7.1 addresses the load imbalance.

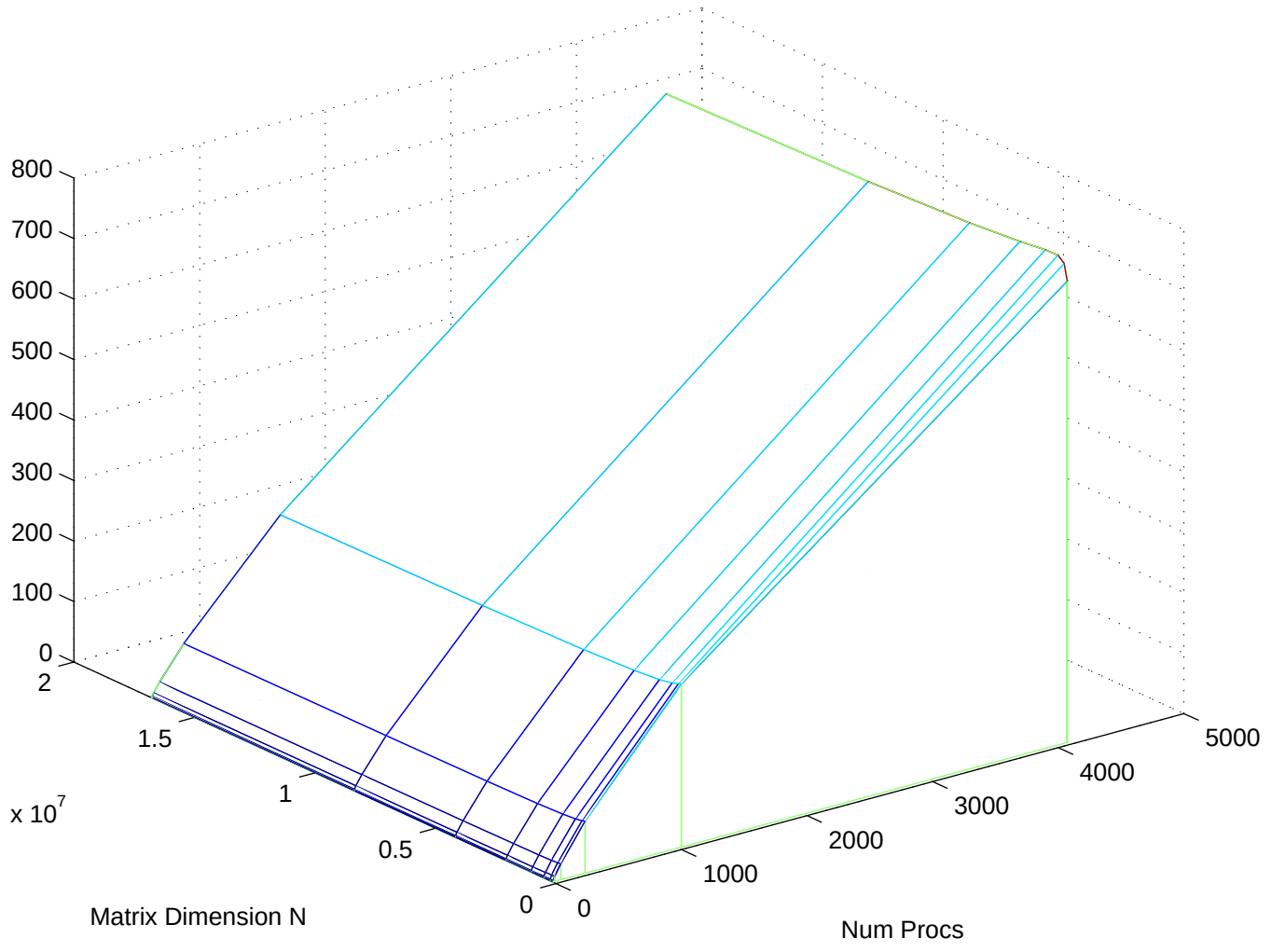

Figure 18: Modeled speedup of synchronous Sparse Cannon

We also evaluate the effects of overlapping communication with computation. Following Krishnan and Nieplocha [39], we define the non-overlapped percentage of communication as:

$$w = 1 - \frac{T_{comp}}{T_{comm}} = \frac{T_{comm} - T_{comp}}{T_{comm}}$$

The speedup of the asynchronous implementation is:

$$S = \frac{W}{T_{comp} + w(T_{comm})}$$

Figure 19 shows the modeled speedup of asynchronous SpCannon assuming truly one-sided communication. For smaller matrices with dimensions $n = 2^{17} - 2^{20}$, speedup is about 25% more than the speedup of the synchronous implementation.

The modeled speedup plots should be interpreted as upper bounds on the speedup that can be achieved on a real system using these algorithms. Achieving these speedups on real systems requires all components to be implemented and working optimally. The conclusion we derive from those plots is that no matter how hard we try, it is impossible to get good speedup with the current 1D algorithms.

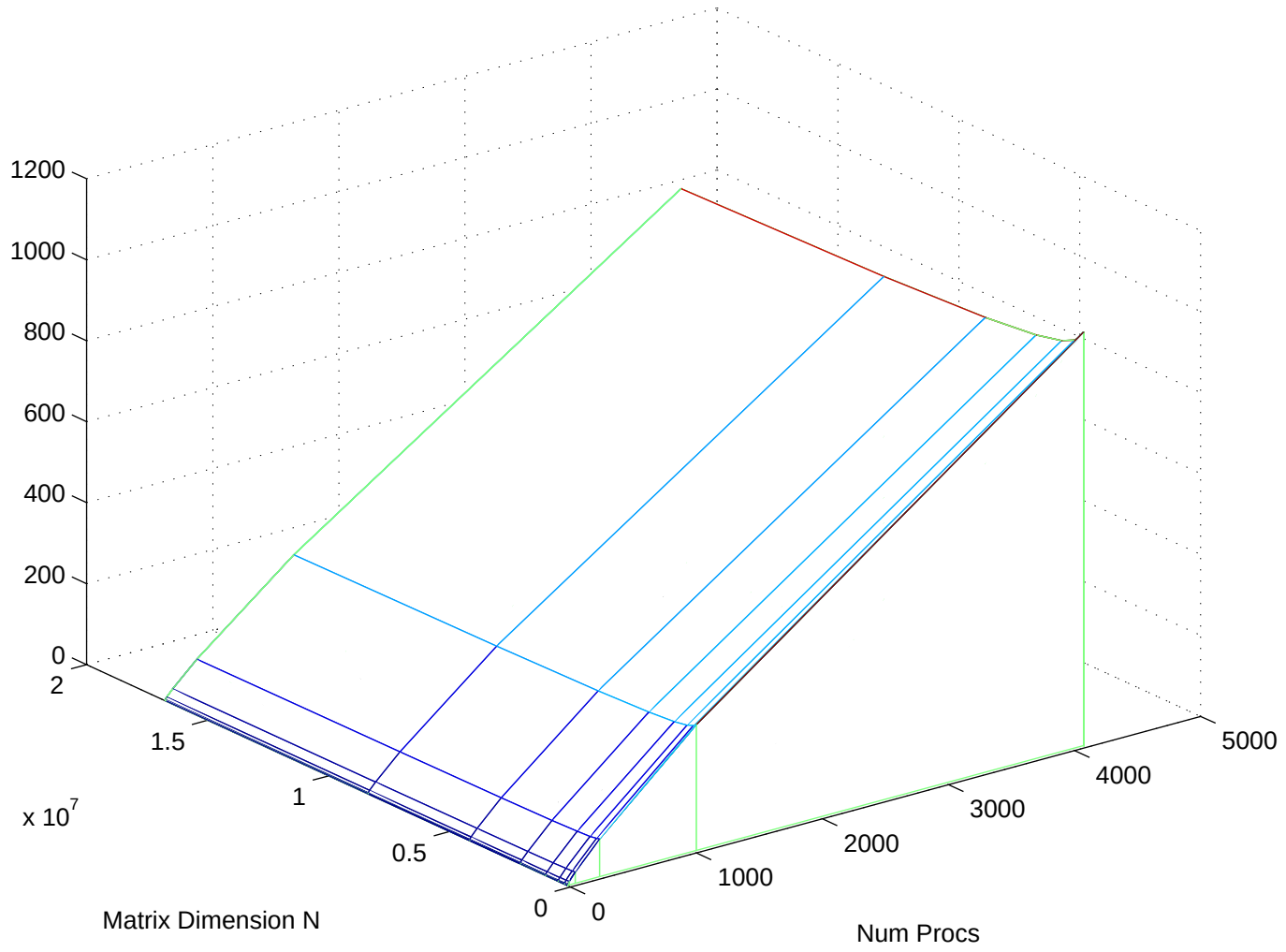

Figure 19: Modeled speedup of asynchronous Sparse Cannon

### *5.3. Scalability with Hypersparsity*

This modeling study reveals the scalability of our hypersparse algorithm with increasing sparsity. We have implemented our data structures and multiplication algorithms in C++. Our code is compiled using the GNU Compiler Collection (GCC) Version 4.1, with the flags `-O3`, because these are the settings that our comparison platform, M , is compiled with. We have incorporated Peter Sander's Sequence Heaps [40] for all the priority queues used by our algorithms. Througout the experiments, the numerical values are represented as double-precision floating points.

We compare the performance of our implementation with M R2007A's (64-bit version) implementation of the classical algorithm. Sparse matrix multiplication is a built-in function in M , so there are no interpretation overheads associated with it. We are simply comparing our C++ code with the underlying precompiled C code used in M .

All of our experiments are performed on a single core of Opteron 2.2 Ghz with 64 GB main memory, where we simulate the execution of a parallel SpGEMM. The simulation is done by dividing the input matrices of size $n \times n$ into $p$ submatrices of size $(n/\sqrt{p}) \times (n/\sqrt{p})$ using the 2D block decomposition, as explained in Section 3.2 and shown in Figure 4.

Expressing the matrix multiplication as algebraic operations on submatrices instead of individual elements, we see that each submatrix of the product is computed using $\mathbf{C}_{ij} = \sum_{k=1}^{\sqrt{p}} \mathbf{A}_{ik}\,\mathbf{B}_{kj}$. Since we are primarily concerned with the sequential sparse matrix multiplication kernel, we will exclude the cost of submatrix additions and other parallel overheads. That is to say, we will only time the submatrix multiplications, exactly plotting

$$time(p, \mathbf{A}, \mathbf{B}) = \sum_{i=1}^{\sqrt{p}} \sum_{j=1}^{\sqrt{p}} \sum_{k=1}^{\sqrt{p}} time(\mathbf{A}_{ik}\,\mathbf{B}_{kj}),$$

which is equal to the amount of work done by a parallel matrix multiplication algorithm such as SUMMA [29].

Increasing $p$ in this case does not mean we use more processors to compute the product. Instead, it means we use smaller and smaller blocks while computing the product on a single processor. Therefore, a perfectly scalable algorithm would yield flat timing curves as $p$ increases. We expect our hypersparse algorithm to outperform the classical algorithm as $p$ increases due to reasons explained in Section 2.2. We label the classical algorithm M , and our algorithm H _GEMM in the plots. The second input is only transposed once because this is what would happen in a parallel implementation.

In all experiments in this section, the input matrices have dimensions $2^{23} \times 2^{23}$, i.e. the input graphs have around 8 million vertices.

### *5.3.1. Synthetic R-MAT Graphs*

We ran two main sets of multiplication experiments with R-MAT matrices, one where both input matrices are R-MAT, and one where **A** is a R-MAT matrix and **B** is a permutation matrix. The results are shown in Figures 20(a) and 20(b).

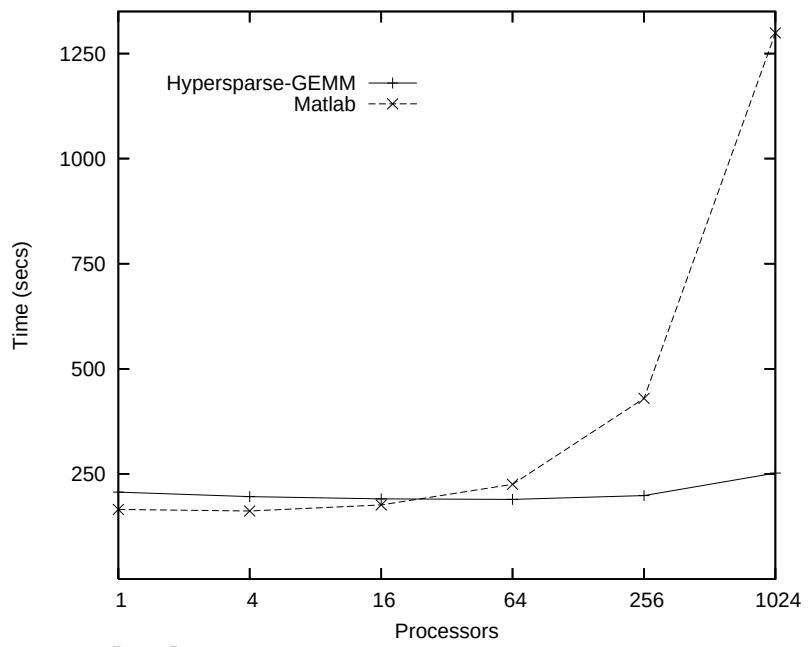

(a) Multiplying R-MAT matrices (R-MAT × R-MAT)

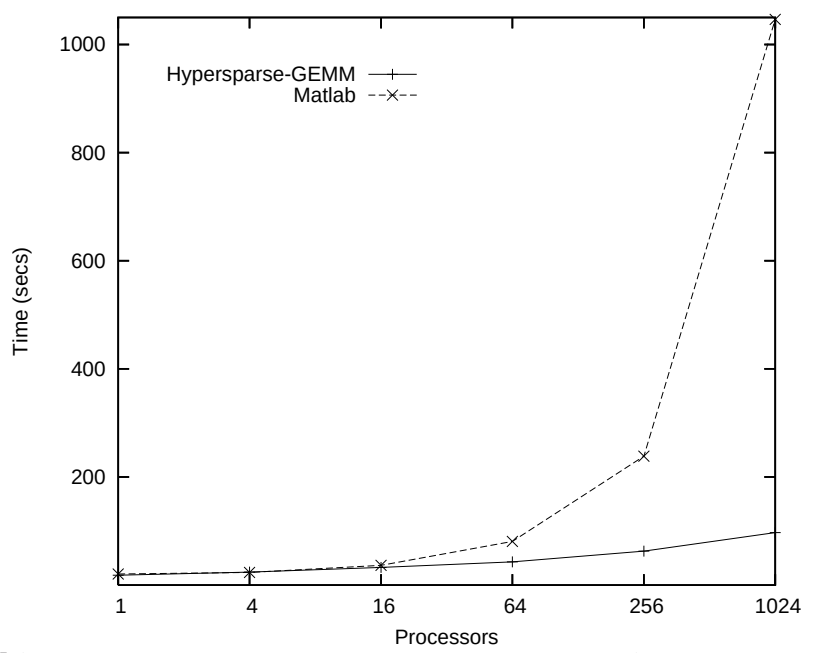

(b) Permuting an R-MAT matrix (R-MAT × Perm)

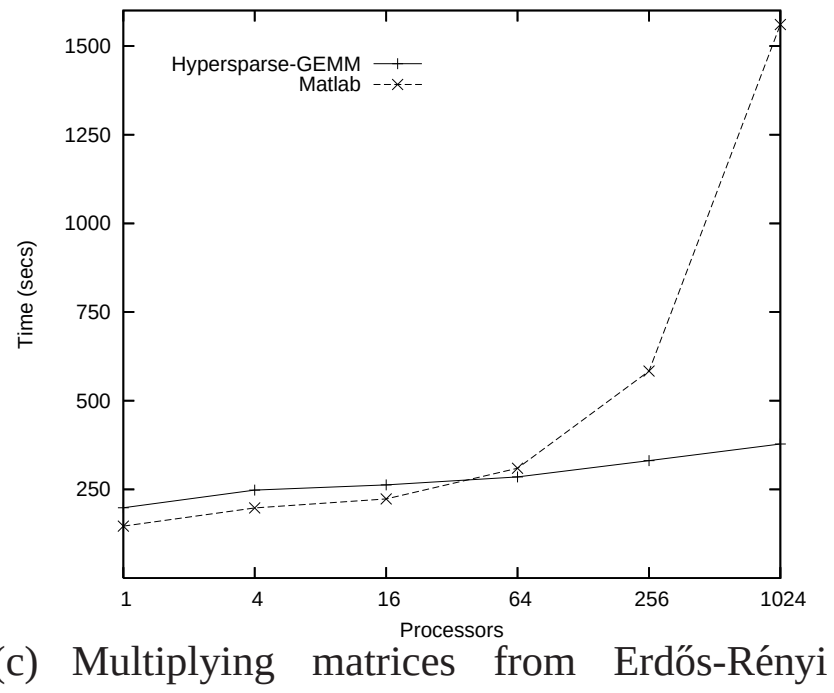

(c) Multiplying matrices from Erdős-Rényi graphs (Rand × Rand)

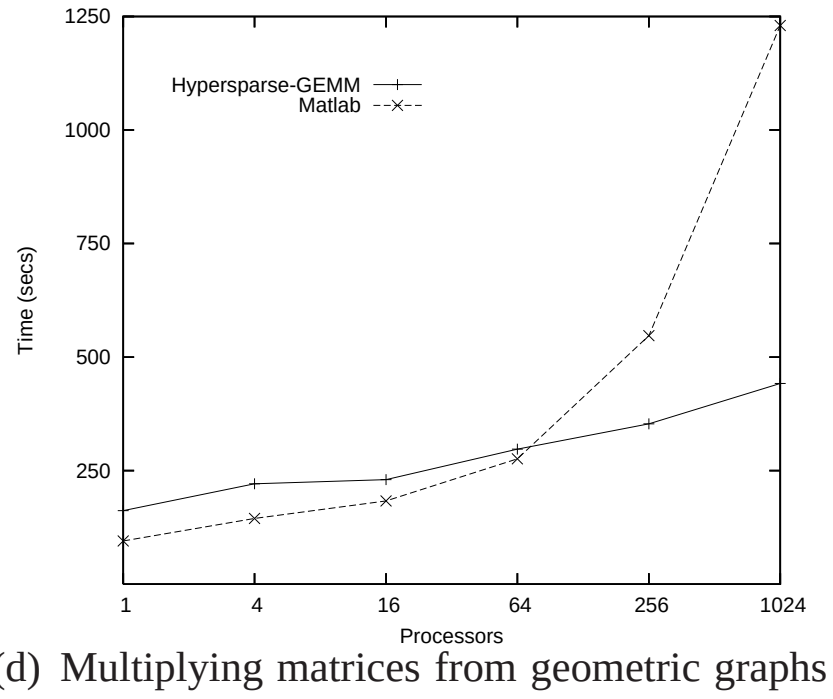

(d) Multiplying matrices from geometric graphs (`grid3d` × `grid3d`)

Figure 20: Model of scalability of SpGEMM kernels

In the case of R-MAT $\times$ R-MAT, the classical sequential algorithm is initially faster than H _GEMM. For $p > 64$, however, the classical algorithm starts performing poorly because submatrices start getting hypersparse. To see why, consider the ratio of *nnz* to $n$ for each submatrix:

$$\frac{nnz(\mathbf{A}_{ij})}{n/\sqrt{p}} = \frac{8\,n/p}{n/\sqrt{p}} = \frac{8}{\sqrt{p}}$$

This ratio is smaller than 1 for $p > 64$, and does to 0 as $p$ increases, making submatrices hypersparse. For $p = 1024$, our algorithm performs more than 5 times faster than the classical algorithm. Its scaling is also very good, showing almost flat curves.

In the case of multiplying an R-MAT matrix with a permutation matrix (R-MAT $\times$ Perm), poor scalability of the classical algorithm is more apparent. Our algorithm starts to outperform for as low as $p > 4$. The break-even point after which our algorithm dominates is lower in this case because permutation matrices are more sparse with only 1 nonzero per column/row.

*5.3.2. Erdős-Rényi Random Graphs*

We have conducted a single set of experiments where we multiply two matrices representing Erdős-Rényi random graphs. Looking at the timings shown in Figure 20(c), we see that the H _GEMM dominates the classical algorithm (as implemented in Matlab) for most values for $p > 64$, when used as the sequential kernel of a 2D parallel SpGEMM. More importantly, when we reach thousands of processors, our algorithms show their scalability for these input types as well. In particular, H _GEMM is more than 4 times faster than the classical algorithm for 1024 processors when multiplying Erdős-Rényi random matrices.

*5.3.3. Regular 3D Grids*

For our last set of experiments, we have used `grid3d` matrices. These matrices have a banded structure, which makes them unsuitable for 2D block decomposition since the off-diagonal processors sit idle without storing any nonzeros and performing any computation. Even though we are just timing the computational costs, ignoring parallelization overheads in this modeling study, the imbalance has an effect on the timing of submatrix multiplications. In particular, the heavy diagonals avoid hypersparsity to emerge, thus favoring the classical algorithm in this unrealistic setting.

To remedy this problem, we perform random permutations of vertices on both inputs before performing the multiplication. In other words, instead of computing $\mathbf{C} = \mathbf{AB}$, we compute $\mathbf{C}' = \mathbf{A}'\mathbf{B}' = (\mathbf{PAP^T})(\mathbf{PBP^T}) = \mathbf{PCP^T}$. Even after applying random symmetric permutations, submatrices in the diagonal are expected to have more nonzeros than others. This is because symmetric permutations essentially relabel the vertices of the underlying graph, so they are unable to scatter the nonzeros in the diagonal.

Multiplications among diagonal blocks favor the classical sequential kernel because diagonal blocks can never become hypersparse no matter how much $p$ increases. Multiplication among off-diagonal blocks are more suitable for our hypersparse kernel. More technically, our observation means

$$\text{flops}(\mathbf{A}_{ii}\,\mathbf{B}_{ii}) > \text{flops}(\mathbf{A}_{ii}\,\mathbf{B}_{ij}) > \text{flops}(\mathbf{A}_{ik}\,\mathbf{B}_{kj}).$$

Therefore, the variances in timings of submatrix multiplications are large compared with other sets of test matrices.

Asymptotic behavior of the algorithms is also slightly different in this case as it can be seen in Figure 20(d). Yet, our algorithm is around 4 times faster than the classical algorithm for $p = 1024$.

## 6. Parallel Scaling of Sparse SUMMA

### *6.1. Experimental Design*

We have implemented two versions of the 2D parallel SpGEMM algorithms in C++. The first one is directly based on Sparse SUMMA and synchronous in nature. It does not use any MPI-2 features. The second implementation is asynchronous and uses one-sided communication features of MPI-2. In this section, we report on the performance of the synchronous implementation only and leave the results of the asynchronous implementation to Section 7.3. We ran our code on the TACC's Ranger Cluster, which has four 2.3GHz quad-core processors in each node (16 cores/node). It has an Infiniband interconnect with 1GB/sec unidirectional point-to-point bandwidth and 2.3 microseconds max latency. We have experimented with multiple compilers and MPI implementations. We report our best results, which we achieved using `OpenMPI v1.3b` and `GNU Compiler (g++ v4.4)` with flag `-O3`.

For both implementations, our sequential H _GEMM routines return a set of intermediate triples that are kept in memory up to a certain threshold without being merged immediately. This allows for a more balanced merging, thus eliminating some unnecessary scans that degraded performance in a preliminary implementation [31].

In our experiments, instead of using random matrices (matrices from Erdős-Rényi random graphs), we used synthetically generated RMAT matrices, in order to achieve results closer to reality. The average number of nonzeros per column is 8 for those synthetically generated graphs.

### *6.2. Experimental Results*

#### *6.2.1. Square Sparse Matrix Multiplication*

In the first set of experiments, we multiply two R-MAT matrices that are structurally similar. This square multiplication is representative of the expansion operation used in the Markov clustering algorithm [41]. It is also a challenging case for our implementation due to high skew in nonzero distribution. We performed strong scaling experiments for different matrix dimensions ranging from $2^{21}$ to $2^{23}$. Figure 21 shows the speedup we achieved. The graph shows linear speedup (with slope 0.5) until around 50 processors; afterwards the speedup is proportional to the square root of the number of processors. Both results are in line with our analysis in Section 4.2.

#### *6.2.2. Tall Skinny Right Hand Side Matrix*

The second set of experiments involves multiplication of R-MAT matrices by tall skinny matrices of varying sparsity. This set of experiments serves multiple purposes. Together with the next set of experiments, they reveal the sensitivity of our algorithm to matrix orientations. It also examines the sensitivity to sparsity, because we vary the sparsity of the right hand side matrix. Lastly, it is representative of the parallel breadth-first search that lies in the heart of our distributed-memory betweenness centrality implementation, which is described in more detail in Chapter 5 of the first author's thesis [17]. We varied the sparsity of the right hand side matrix from approximately 1 nonzero per column to $10^5$ nonzeros per column, with multiplicative increments of 10. This way, we imitate the patterns of the betweenness centrality execution where at each level of breadth-first search, the current frontier (fringe) has as low as a few vertices but it can have as high as 300000 vertices. Figure 22 plots the number of vertices in the fringe at each level of the breadth-first search for 10 different runs (with different starting vertices) on a network of 1 million vertices and 8 million edges.

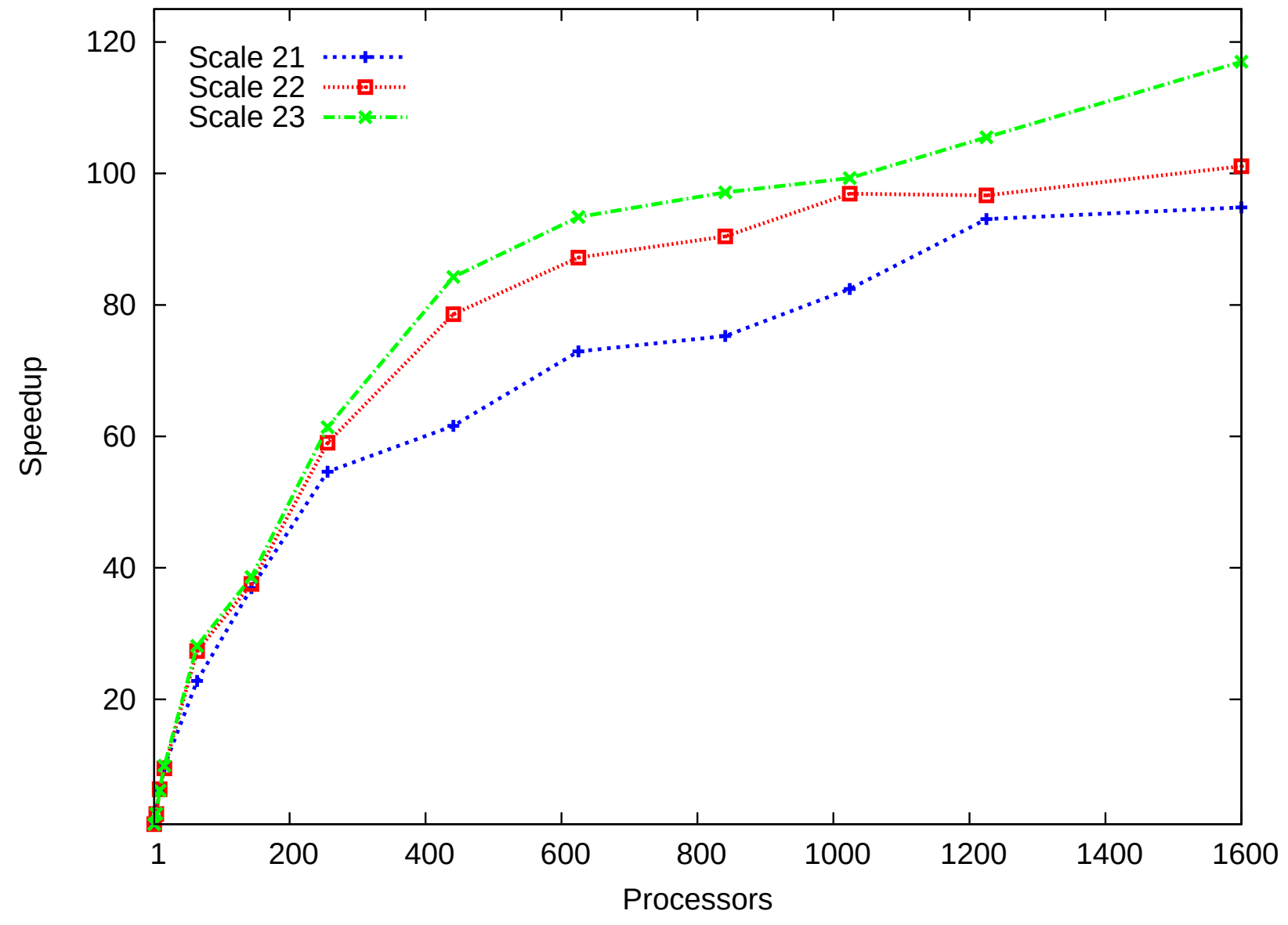

Figure 21: Observed speedup of of synchronous Sparse SUMMA for the R-MAT $\times$ R-MAT product on matrices having dimensions $2^{21} - 2^{23}$. Both axes are normal scale.

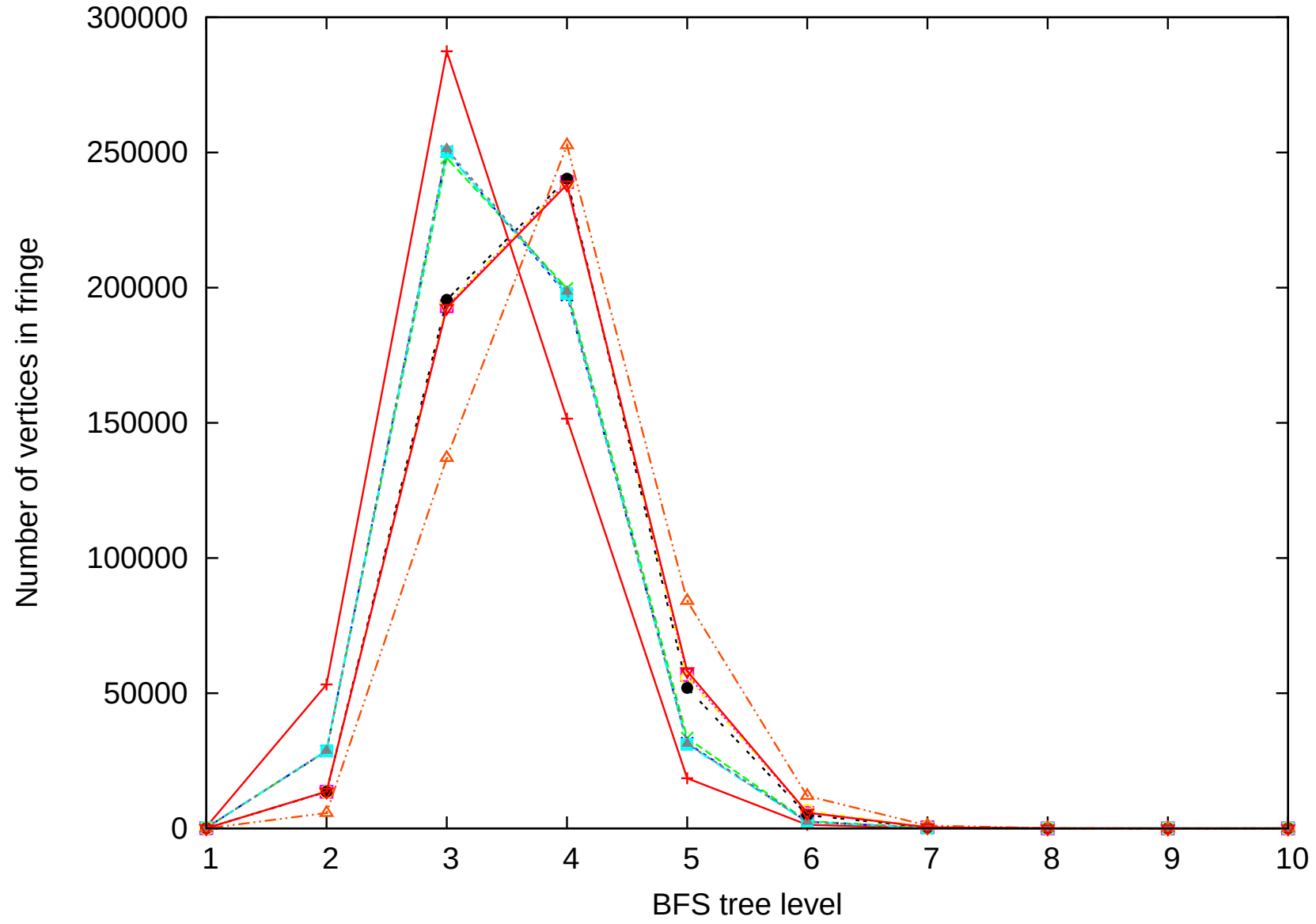

Figure 22: Fringe size per level during breadth-first search. Each one of ten plots is an average of 256 independent BFS operations on a graph of 1 million vertices and 8 million edges

For our experiments, the R-MAT matrices on the left hand side have $c_1 = 8$ nonzeros per column and their dimensions vary from $n = 2^{20}$ to $n = 2^{26}$. The right hand side matrix is of size $n$-by-$k$, and its number of nonzeros per column, $c_2$ is varied from 1 to $10^5$, with multiplicative increments of 10. Its width, $k$, varies from 128 to 8192 that grows proportionally to its length $n$. Hence, the total work is $W = O(c_1 c_2 k)$, the total memory consumption is $M = O(c_1 n + c_2 k)$, and total bandwidth requirement is is $O(M\sqrt{p})$.

We performed scaled speedup experiments where keep both $n/p = 2^{14}$ and $k/p = 2$ constant. This way, we were able to keep both memory consumption per processor and work per processor constant at the same time. However, bandwidth requirements per processor increases by a factor of $\sqrt{p}$.

Figure 23 shows the three-dimensional performance graph. The timings for each slice along the XZ-plane (i.e. for every $c_2 = \{1, 10, ..., 10^5\}$ contour), is normalized to its running time on $p = 64$ processors. We do not cross-compare the absolute performances using different $c_2$ values, as our focus in this section is parallel scaling. The graph demonstrates that, except for the outlier case $c_2 = 1000$, we achieve the expected $\sqrt{p}$ slowdown due to communication costs. The performance we achieved for these large scale experiments, where we ran our code on up to 4096 processors, is remarkable.

#### *6.2.3. Multiplication with the Restriction Operator*

The multilevel method is widely used in the solution of numerical and combinatorial problems [42]. The method constructs smaller problems by successive coarsening of the problem domain. The simplest coarsening is perhaps graph contraction. One contraction step chooses two or more vertices in the original graph $G$ to become a single aggregate vertex in the contracted graph $G'$. The edges of $G$ that used to be incident to any of the vertices forming the aggregate now become incident to the new aggregate vertex in $G'$.

Constructing a coarser grid during the V-cycle of the Algebraic Multigrid (AMG) method [11] or graph partitioning [43] is a generalized graph contraction operation. Different algorithms need different coarsening operators. For example, a weighted (as opposed to strict) aggregation [44] might be preferred for partitioning problems. In general, coarsening can be represented as multiplication of the matrix representing the original fine domain (grid, graph, or hypergraph) by the restriction operator.

In this experiments, we use a simple restriction operation to perform graph contraction. Gilbert et al. [6] describe how to perform contraction using SpGEMM. Their elegant algorithm creates a special sparse matrix $\mathbf{S}$ with $n$ nonzeros. The triple product $\mathbf{SAS}^\mathsf{T}$ contracts the whole graph at once. Making $\mathbf{S}$ smaller in the first dimension while keeping the number of nonzeros same changes the restriction order. For example, we contract the graph into half by using $\mathbf{S}$ having dimensions $n/2 \times n$, which is said to be of order 2.

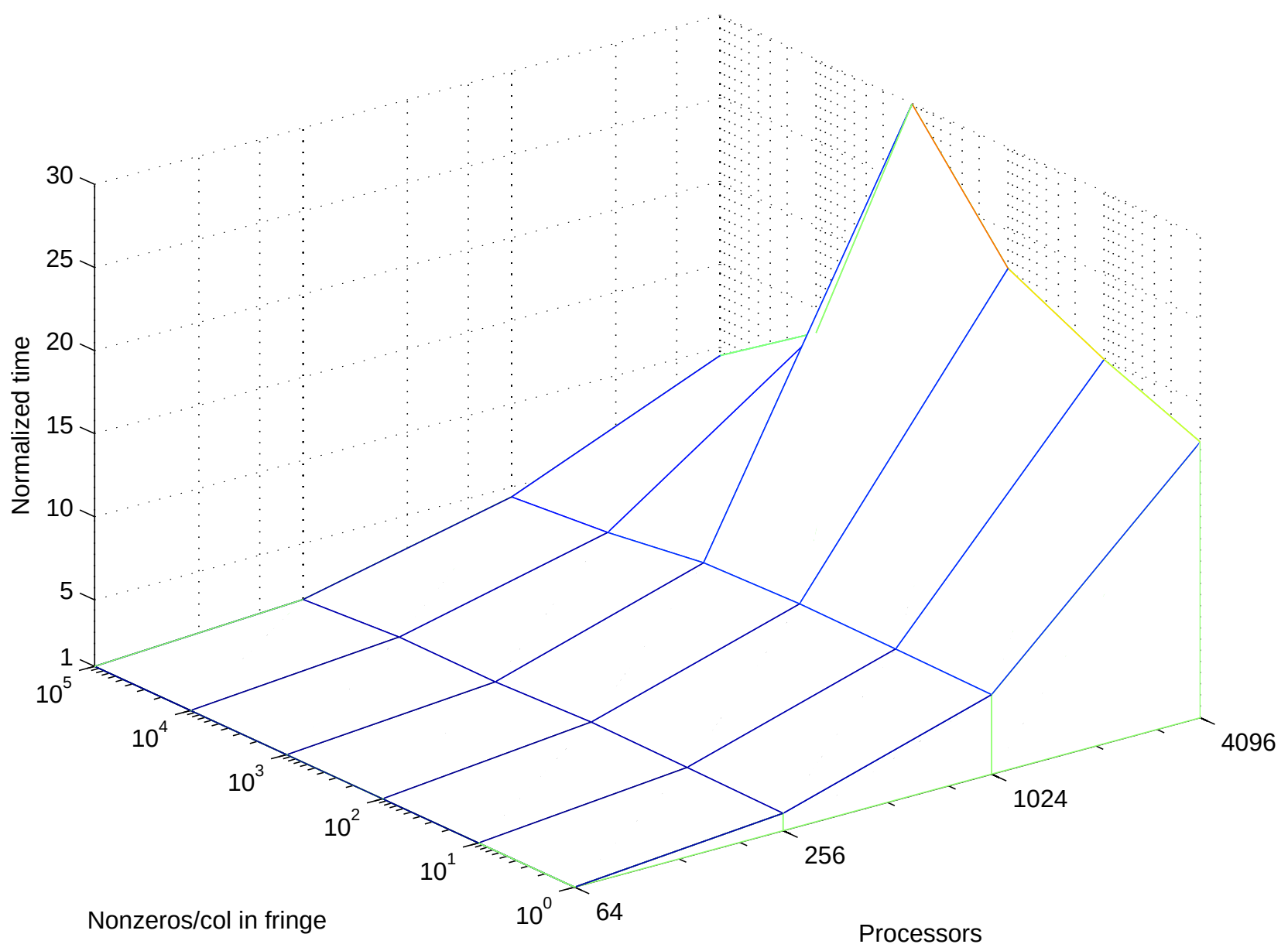

Figure 23: Weak scaling of R-MAT times a tall skinny Erdős-Rényi matrix. x (processors) and y (nonzeros per column on fringe) axes are logarithmic, whereas z (normalized time) axis is normal scale.

Figure 24 shows the strong scaling of the operation $\mathbf{AS}^{\mathsf{T}}$ for R-MAT graphs of scale 23. We used restrictions of order 2, 4, and 8. Changing the interpolation order results in minor changes in performance. The experiment shows good scaling for up to 1024 processors.

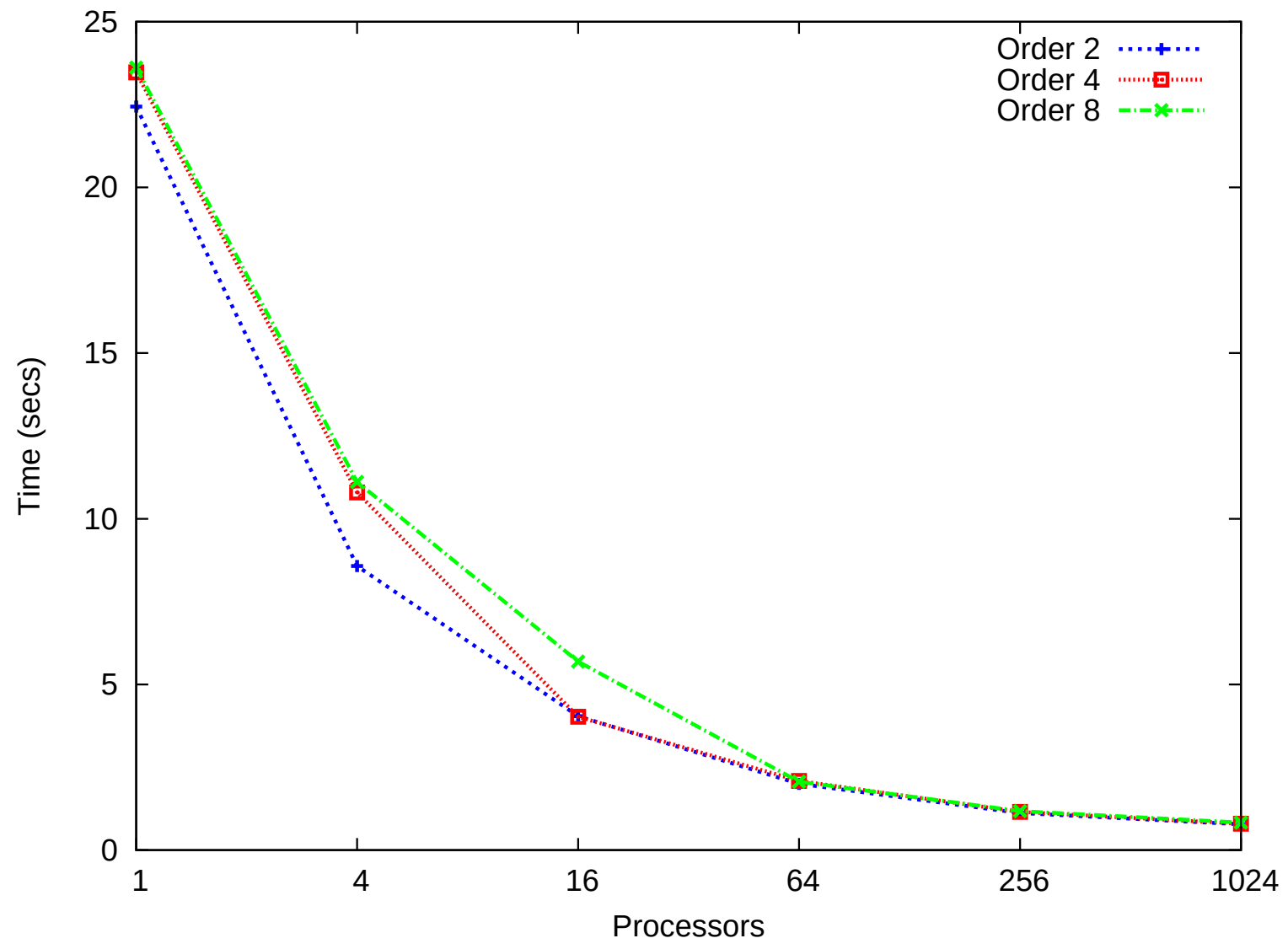

Figure 24: Strong scaling of multiplication with the restriction operator on the right, $\mathbf{A}' \leftarrow \mathbf{A}\mathbf{S}^{\mathsf{T}}$. The graph is logarithmic on the x-axis.

## 7. Alternative Parallel Approaches

### *7.1. Load Balancing and Asynchronous Algorithms*

In distributed memory dense matrix-matrix multiplication algorithms, each processor performs a total of $W/p$ work where $W = N^3$. The sparse inputs are not so naturally balanced. Our experiments with randomly relabeling vertices (in matrix terms, applying a symmetric permutation) showed good premise where the maximum overall work for a single processor was only 9% more than the average work per processor, even when the initial matrix has significantly skewed degree distribution[2]. Aiming for perfect load balance via graph or hypergraph partitioning [45, 46, 47] seems impractical whenever the matrices are not reused. Even when one of the matrices are fixed throughout the computation, load balance for SpGEMM can not be determined solely based on one operand, unlike SpMV. We do not know of any applications where both matrix operands have fixed structure for several subsequent multiplication operations, which might have justified complex load balancing.

The sparse 2D algorithms presented in previous sections execute in a synchronous manner in $s$ stages in their naive form. For sparse matrices, achieving good load balance per stage is harder than achieving load balance for the whole computation. This is because a local submatrix update such as $\mathbf{C}_{i,j} \leftarrow \mathbf{C}_{i,j} + \mathbf{A}_{i,k}\mathbf{B}_{k,j}$ might have significantly more work to do than another update at the same stage, say $\mathbf{C}_{i+1,j} \leftarrow \mathbf{C}_{i+1,j} + \mathbf{A}_{i+1,k}\mathbf{B}_{k,j}$. However, in a subsequent stage the roles of the $(i, j)$th and the $(i+1, j)$th processor might swap; hence balancing the load across stages. On the other hand, a barrier synchronization at each stage forces everyone to wait for the slowest update until they can proceed to the next stage. Hence, we expect an asynchronous algorithm to perform better than a synchronous one for matrices with highly skewed nonzero distribution.

In order to quantify the severity of load imbalance, we performed a simulation of the Sparse Cannon algorithm that accounts for the computation (in terms of the number of actual flops only) and communication (*nnz* only) done by each processor. We varied the matrix dimension and the number of processors while the number of nonzeros per row/column were kept constant. For RMAT matrices with 8 nonzeros per column, the per-stage load imbalance with 256 processors is shown in Figure 25. Load imbalance is defined as the ratio of the maximum number of flops performed by any processor to the average number of flops. These plots are typical in the sense that we permuted the input matrices multiple times with different random permutations and plotted the results of the permutation that resulted in the median load imbalance.

Figure 26(a) shows the overall load imbalance for increasing matrix sizes on 256 processors. The problem becomes well balanced (i.e. it has less 10% load imbalance) for R-MAT inputs of scale 20 and larger. On the other hand, Figure 26(b) shows a comparison of the trends of overall and per-stage imbalances (average over all stages) with increasing number of processors and a fixed problem size.

These results on Figures 25 and 26 suggest that per-stage load balance is significantly harder to achieve than load balance for the overall computation. Both tend to decrease as the problem size gets bigger, although per-stage load imbalance has much

---

[2]For sufficiently large matrices on 256 processors, as shown in Figure 26

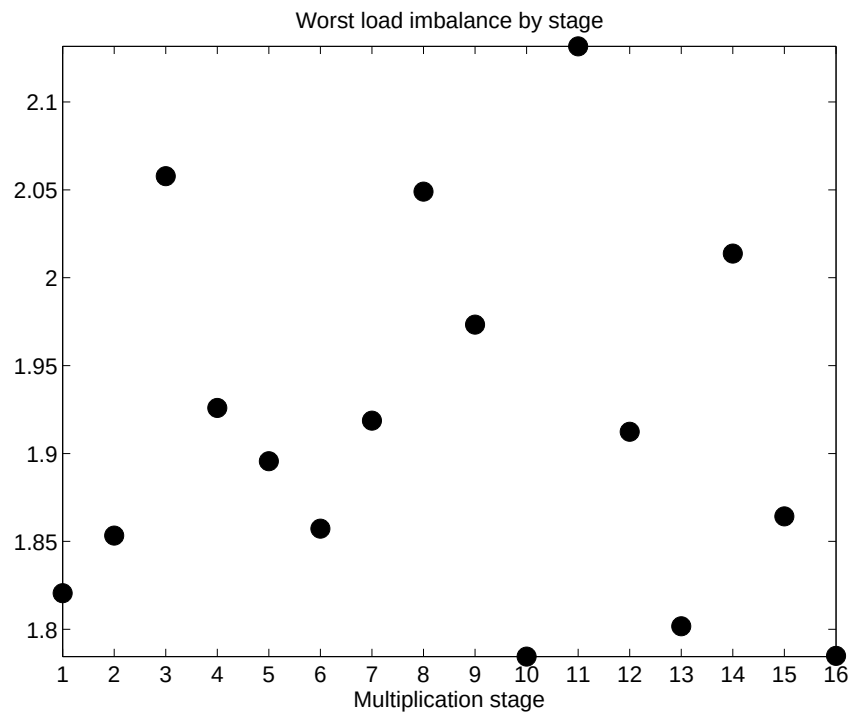

(a) RMAT Scale-15

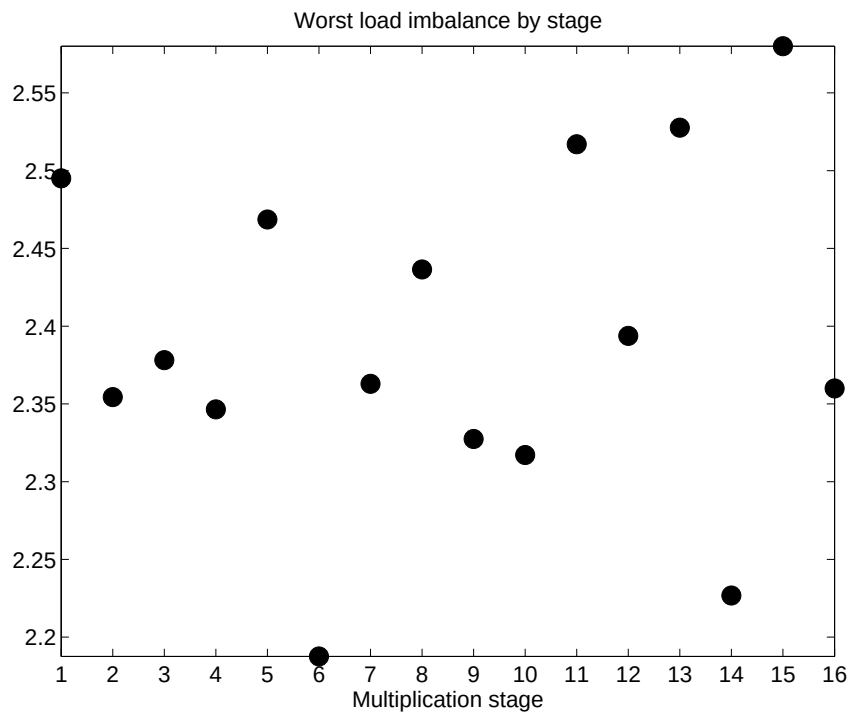

(b) RMAT Scale-16

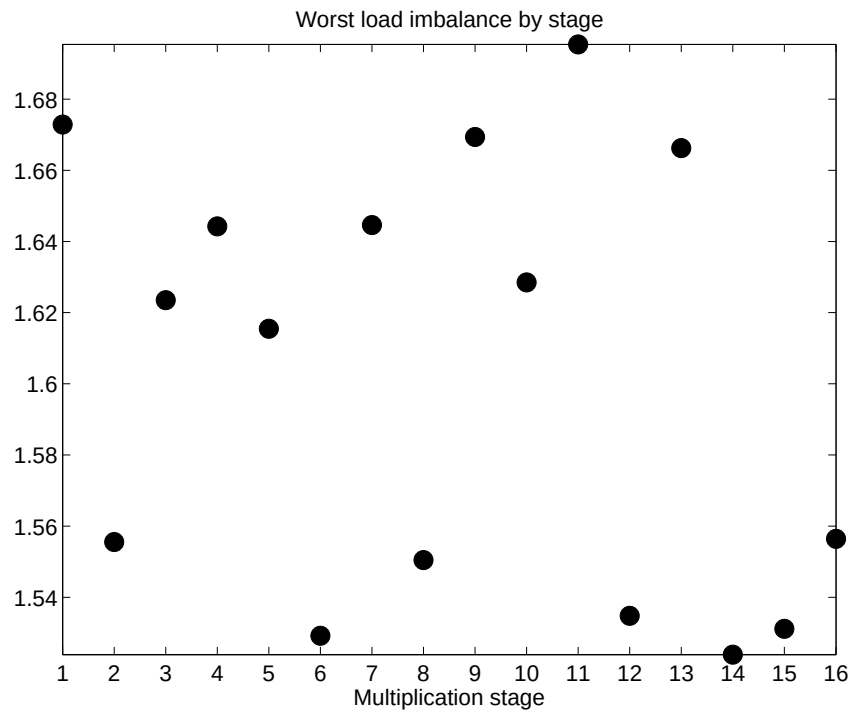

(c) RMAT Scale-17

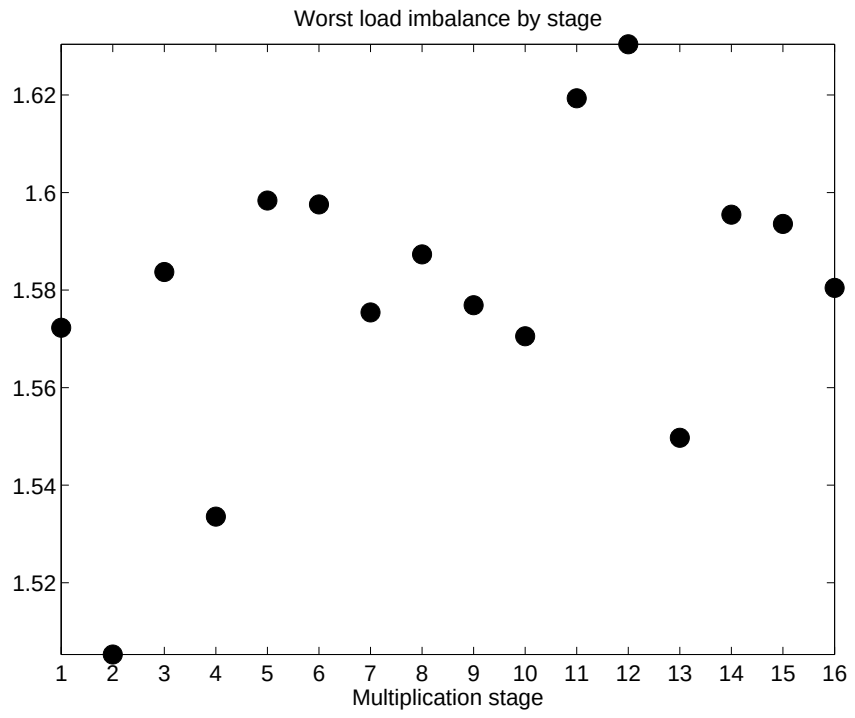

(d) RMAT Scale-18

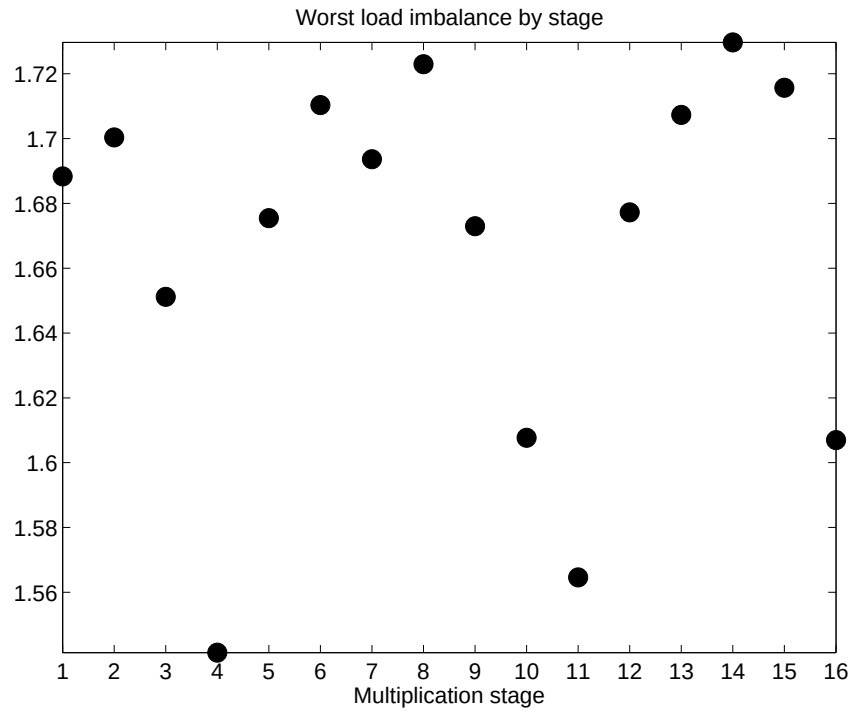

(e) RMAT Scale-19

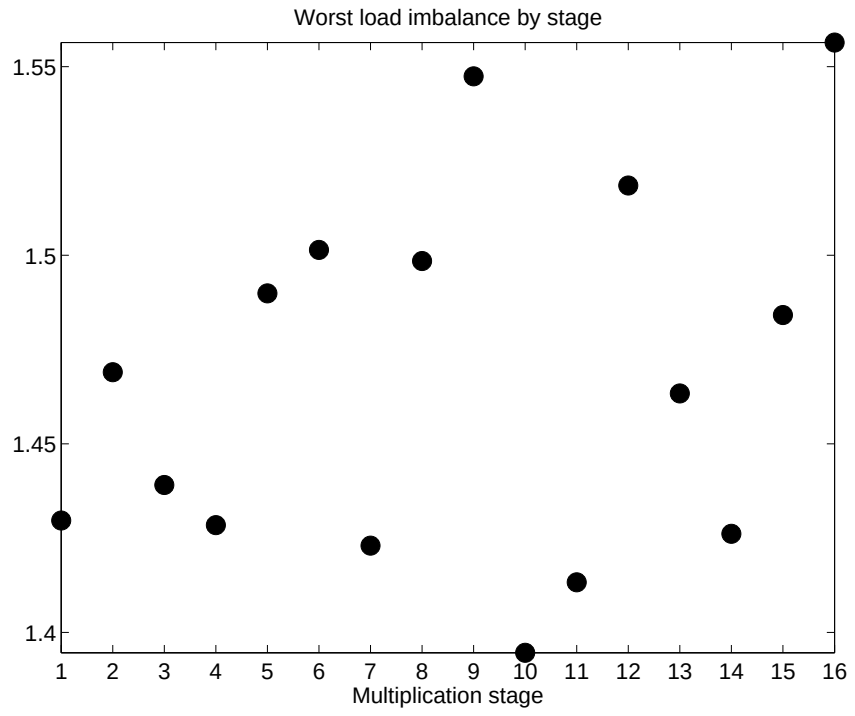

(f) RMAT Scale-20

Figure 25: Load imbalance per stage for multiplying two RMAT matrices on 256 processors using Sparse Cannon

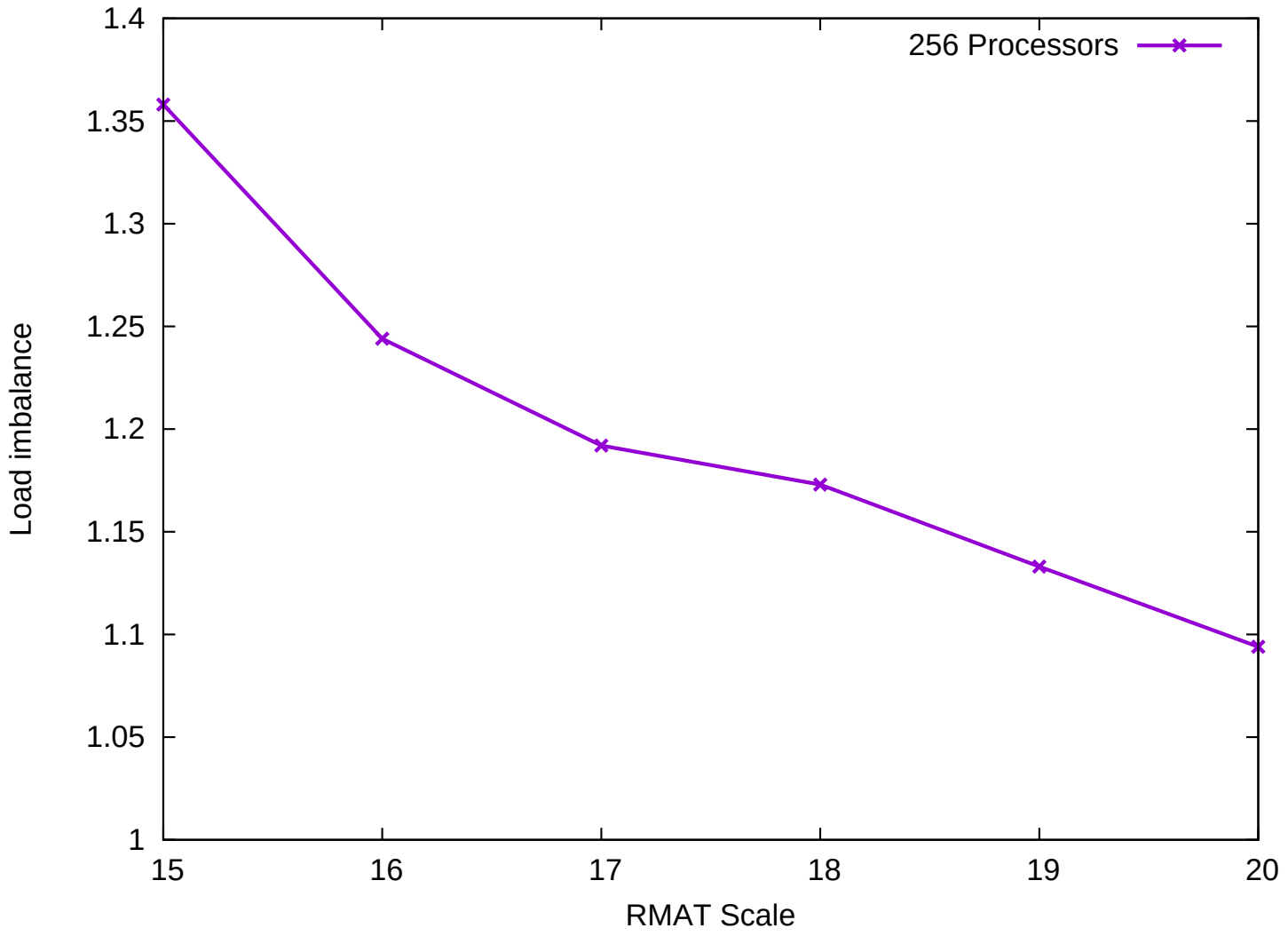

(a) Overall imbalance on a fixed number of ($p$ = 256) processors

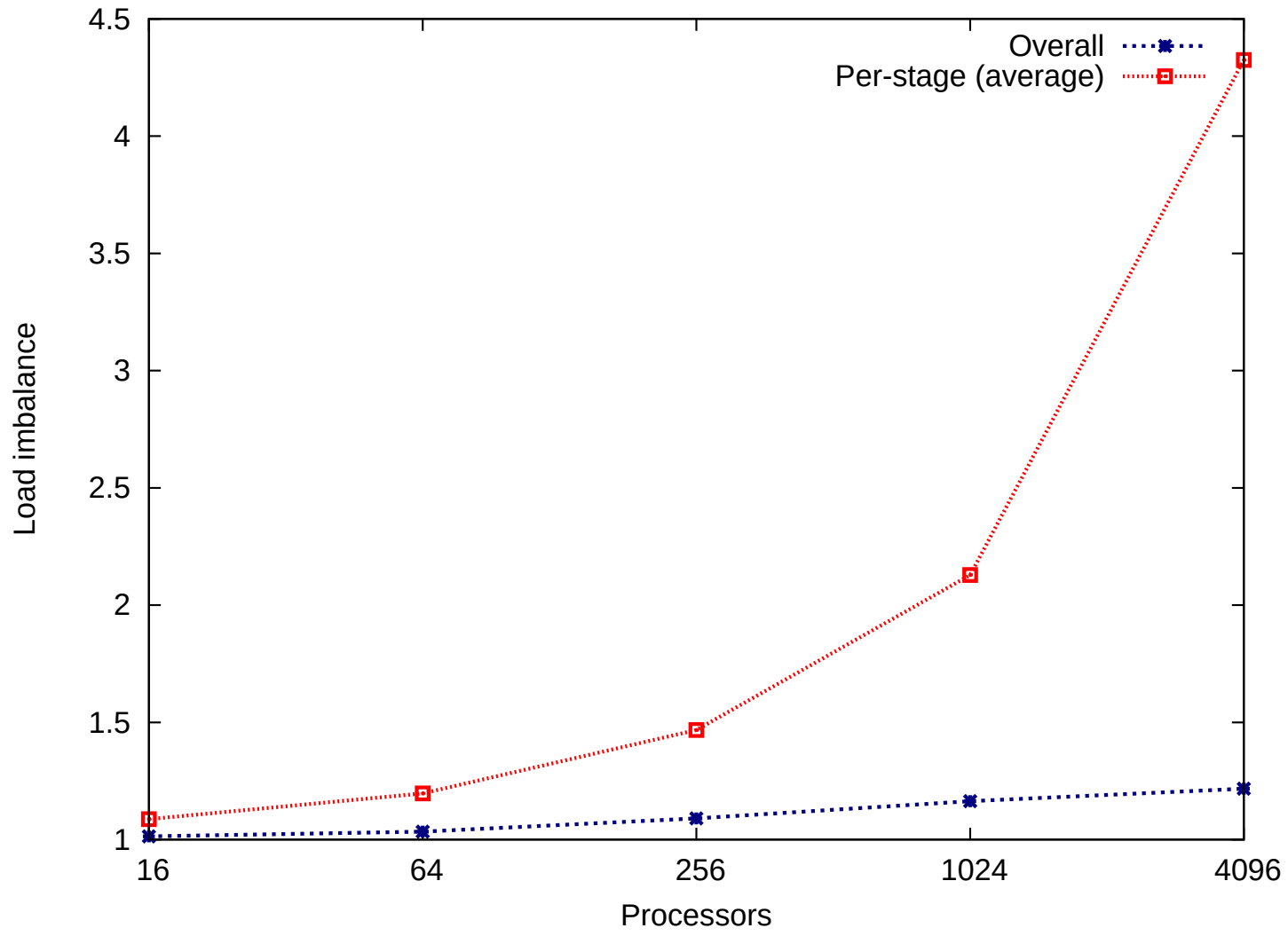

(b) Overall vs. per-stage imbalance for a fixed problem size (R-MAT scale 20)

Figure 26: Load imbalance during parallel multiplication of two RMAT matrices

wider variance and tends to decrease less smoothly than the overall load imbalance. The average per-stage load imbalance across all stages is 1.46 for inputs of scale 20. This means that a synchronous Sparse Cannon is likely to achieve 46% less speedup than we estimated in Section 5.2. By contrast, a perfectly asynchronous implementation would only pay 9% performance penalty due to load imbalance.

One-sided communication is the most suitable paradigm for implementing asynchronous SpGEMM. We used one-sided MPI-2 routines for portability, as GASNet [48] and ARMCI [49] are not as widely supported on supercomputers. It is still worth mentioning that even MPI poses some complications due to immaturity of implementations and vagueness in parts of the standard. We report our performance results using the passive target synchronization [50].

MPI-1 standard is inadequate to address the asynchronous implementation challenge. The blocking operations do trivially synchronize, and the non-blocking operations buffer the message and revert to a synchronous mode whenever the data is too large to fit in the buffers [51]. The basic requirement of an asynchronous SpGEMM is that the $(i, j)$th processor should be able to fetch its required submatrix from its original owner regardless of its computation stage at that moment. Although this can be achieved by the use of a helper thread that waits on the Send() operation, ready to serve any incoming Recv() requests, this approach has two drawbacks. Firstly, there is a substantial performance loss due to oversubscribing the processor. Secondly, general multithreaded MPI support is still in its infancy[3].

### 7.2. Overlapping Communication with Computation

In order to hide communication costs as much as possible, each processor starts prefetching one submatrix ahead while computing its current submatrix product. More concretely, processor $P(i, j)$ starts prefetching $\mathbf{A}_{i,k+1}$ and $\mathbf{B}_{k+1,j}$ while computing $\mathbf{A}_{i,k}\mathbf{B}_{k,j}$. To keep the memory footprint the same as the synchronous Sparse SUMMA, we split the submatrices in half, so that each processor performs $2\sqrt{p}$ submatrix multiply-adds instead of $\sqrt{p}$. The distribution of matrix $\mathbf{A}$ on a single processor row is shown in Figure 27.

### 7.3. Performance of the Asynchronous Implementation

The pseudocode for our asynchronous implementation (in MPI/C++ notation) is shown in Figure 28. This implementation achieves two goals at once. It overlaps communication with computation as much as possible by prefetching subsequent submatrices while working on the multiplication of the current submatrices. It also achieves better load balance because it allows each processor to proceed independently without any global synchronizations.

Figure 29 compares the performance of the asynchronous implementation with the synchronous Sparse SUMMA implementation for the scale 22 R-MAT × R-MAT product. Although they scale similarly well, the synchronous implementation is 6 – 47% faster.

[3]OpenMPI's MPI_THREAD_MULTIPLE support, which failed in our tests, is known to be untested

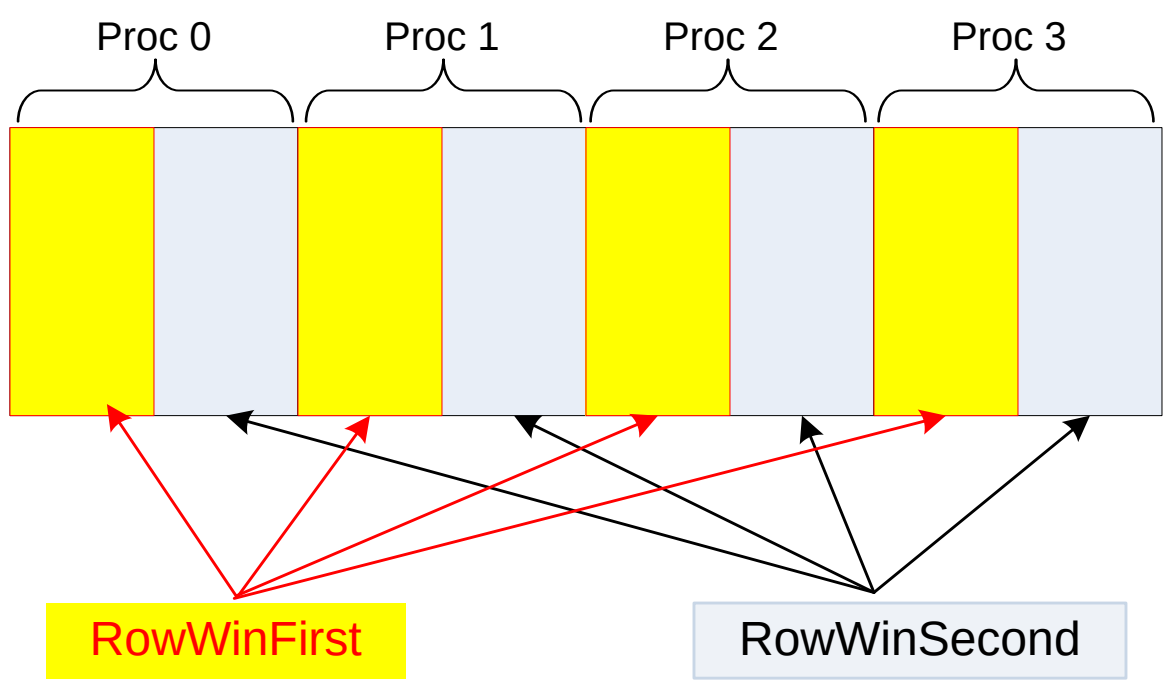

Figure 27: The split distribution of matrix **A** on a single processor row

Overall poor performance of the asynchronous implementation is partly due to the extra operations such as splitting and joining matrices. However, their share in the computation time goes down as we increase the number of processors, so this does not explain the performance difference on large number of processors.

We first thought the performance hit was due to the progress threads that are used by MPI implementations on Infiniband [52] to ensure asynchronous progress. On the other hand, we ran the same code using 4 threads per node so that the progress threads will not oversubscribe the individual cores. Figure 30 shows that the performance difference between the synchronous and asynchronous implementations grows as we use less cores per node. Either our asynchronous implementation, which uses one-sided point-to-point communication instead of blocking collective communication, or the underlying MPI implementation does not take full advantage of the extra bandwidth available per core.

We are not able to explain the load imbalance that happens in practice. Let $T_p$ be the time to complete the SpGEMM procedure on p processors. If $T_i$ is the time for the $i$th processor to complete its local procedure, then $T_p = \max(T_i)$ over all $i$ due to wait times. For the asychronous implementation, our preliminary profiling (on 256 cores) revealed that the fastest processor spends more time waiting for the other processors than doing useful computation. On average, a processor spent about 1/3rd of its time waiting.

The slowdown due to the asynchronous execution was previously experienced on the Connection Machine CM5 [53] on programs with regular communication patterns. Brewer and Kuszmaul [54] found out that an initial skew of processors slowed down the overall computation on the CM5, as receiver queues started to back off. The CM-5 data network is similar to Ranger's, in the sense that they both use a fat-tree [55] interconnect. However, the problem with the CM-5 was the contention on the receivers due to the computational cost of receiving packets. Ranger's Infiniband interconnect, on the other hand, has RDMA support for this task. However, we do not know whether MPI-2 functions have been implemented to fully take advantage of the network's capabilities. In conclusion, revealing the exact cause of the poorer performance of the asynchronous implementation needs further research and more performance profiling.

```
// M1 is the first half of the local matrix M, M2 is the second
vector<Win> rwf = CreateWindows(RowWorld, A1);
vector<Win> rws = CreateWindows(RowWorld, A2);
vector<Win> cwf = CreateWindows(ColWorld, A1);
vector<Win> cws = CreateWindows(ColWorld, A2);

// Each window is made accessible to its neighbors in their
// respective processor row (in the case of A) and
// processor column (in the case of B)
ExposeWindows();

/* Perform initial two fetches and multiply first halfs */

for(int i = 1; i < stages; ++i)     // main loop
{
      CResult += SpGEMM(*ARecv1, *BRecv1, false, true);

      // wait for the previous second halfs to complete
      CompleteFetch(rws);
      CompleteFetch(cws);

      Aowner = (i+Aoffset) % stages;
      Bowner = (i+Boffset) % stages;

      // start fetching the current first half
      StartFetch(ARecv1, Aowner, rwf);
      StartFetch(BRecv1, Bowner, cwf);

      // while multiplying (completed) previous second halfs
      CResult = SpGEMM(*ARecv2, *BRecv2, false, true);

      /* now wait for the current first half to complete */
      /* start prefetching the current second half */
}
/* perform the last pieces of computation */
```

Figure 28: Partial C++ code partial for asynchronous SpGEMM using one-sided communication and split prefetching for overlapping communication with computation

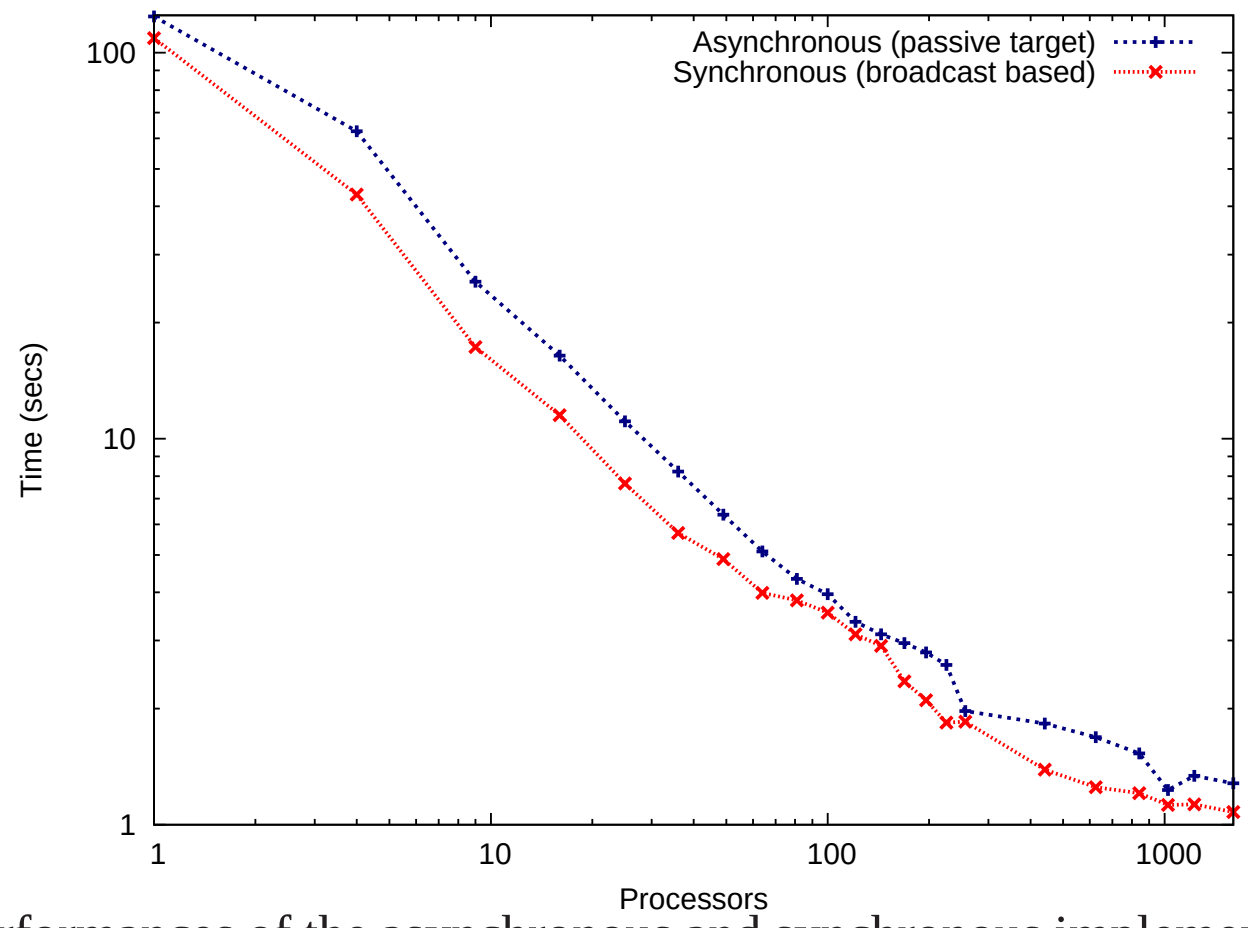

Figure 29: Performances of the asynchronous and synchronous implementations of the Sparse SUMMA. In this experiment, we multiply two R-MAT matrices of scale 22. Both axes are on log-scale

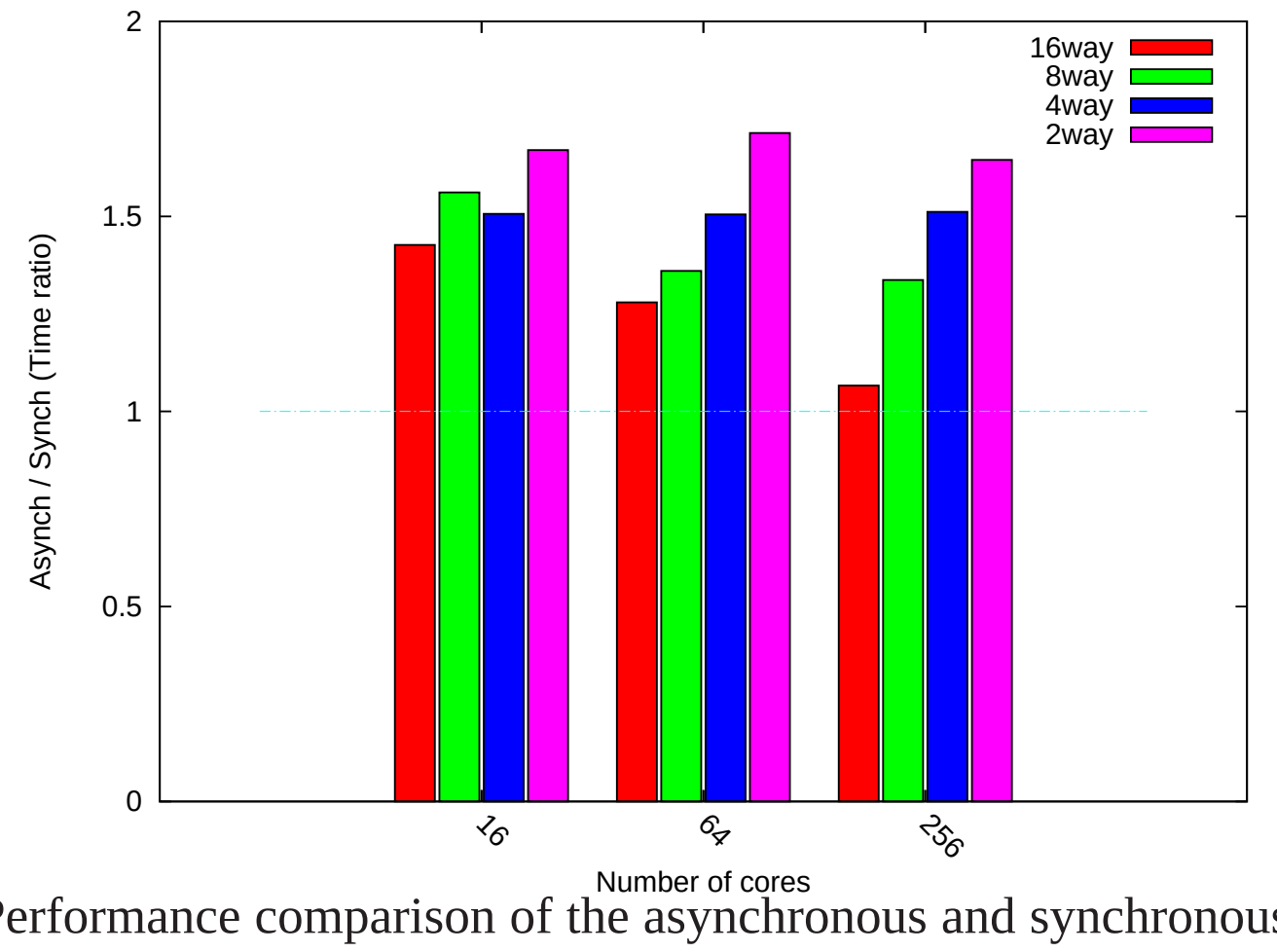

Figure 30: Performance comparison of the asynchronous and synchronous implementations using different number of cores per node. The vertical axis shows the ratio of time it takes to execute SpGEMM with the asynchronous implementation to the time it takes with the synchronous implementation. A t-way run on $p$ cores is performed using $p/t$ nodes.

## 8. Future Work

Our mathematical modeling of the parallel algorithms in Section 3 is an average-case analysis assuming independent uniform random distribution of nonzeros, which translates into the Erdős-Rényi random graph model. More realistic models should assume skewed nonzero distributions, such as power-law distributions. Ultimately, average case analysis has its limitations because it needs to assume an underlying distribution. On the other hand, worst case analysis does not make a lot of sense for our problem, because there are certain sparse matrix pairs that will create a dense output when multiplied. Therefore, a smoothed analysis [56] of the sparse matrix multiplication algorithms, both sequentially and in parallel, would be a significant advancement, although it is far from clear how to apply the principles of smoothed analysis to an algorithm with discrete inputs.

Load imbalance is not severe for sufficiently large matrices, even in the absence of asynchronous progress. Our one-sided communication approach was based on remote get operations in order to avoid fence synchronization. Given the acceptable load balance for large matrices, it is worth exploring an option with fence synchronization and remote put operations. This proposed implementation will still use one-sided communication but all processors in the processor row/column will need to synchronize after the put operation. We expect better performance because it only takes one trip to complete a remote put operation whereas remote get requires a roundtrip.

Our SpGEMM routine might be extended to handle matrix chain products. In particular, the sparse matrix triple product (RAP) is heavily used in the coarsening phase of the algebraic multigrid method [57]. Sparse matrix indexing and parallel graph contraction also require sparse matrix triple product [6]. The support for sparse matrix chain products eliminates temporary intermediate products and allows more optimizations, such as performing structure prediction [25] and finding the optimal parenthesization based on the sparsity of the inputs.

Finally, there is a need for hierarchical parallelism due to vast differences in the costs of inter-node and intra-node communication. The flat parallelism model does not only lose the opportunity to exploit the faster on-chip network, but it also increases the contention on the off-chip links. We observed that the inter-node communication becomes slower as the number of cores per node increases because more processes are competing for the same network link. According to our preliminary experiments on 1024 cores, Sparse GEMM runs more than 80% faster if we use only 4 cores per node, compared to utilizing all 16 available cores per node. Therefore, designing a hierarchically parallel Sparse GEMM algorithm is an important future direction.